\documentclass[twocolumn,tighten]{aastex61}

\usepackage{verbatim} 
\usepackage{natbib}
\usepackage{amsmath}
\usepackage{gensymb}
\usepackage{graphicx}
\usepackage[caption=false]{subfig}
\usepackage{hyperref}
\hypersetup{
    colorlinks=true,
    linkcolor=blue,
    citecolor=blue,
    filecolor=green,
    urlcolor=magenta,
}

\makeatletter
\renewcommand \paragraph{\@startsection{paragraph}{4}{\z@}%
 {2ex plus 1ex minus .2ex}{1ex plus .2ex}{\small\itshape \center}}
\makeatother
\shorttitle{Machine Learning to Sift Through the LSST Alert Stream}
\shortauthors{Narayan, Zaidi, Soraisam et al.}

\begin{document}

\title{Machine Learning-based Brokers for Real-time Classification of the LSST Alert Stream}

\author{Gautham Narayan}
\affiliation{Space Telescope Science Institute, 3700 San Martin Dr., Baltimore, MD 21218, USA}
\altaffiliation{Lasker Fellow}

\author{Tayeb Zaidi}
\affiliation{Macalester College, Department of Physics and Astronomy, 1600 Grand Avenue, Saint Paul, MN 55105, USA}

\author{Monika D. Soraisam}
\affiliation{National Optical Astronomy Observatory, 950 N. Cherry Ave., Tucson, AZ 85719, USA}
\author{Zhe Wang}
\affiliation{University of Arizona, Department of Computer Science, 1040 E. 4th St., Tucson, AZ 85721, USA}

\author{Michelle Lochner}
\affiliation{African Institute for Mathematical Sciences, 6 Melrose Rd., Muizenberg 7945, Cape Town, South Africa}
\affiliation{SKA South Africa, 3rd Floor, The Park, Park Rd., Pinelands, 7405, South Africa}
\affiliation{Department of Physics and Astronomy, University College London,
Gower St., London WC1E 6BT, UK}

\author{Thomas Matheson}
\affiliation{National Optical Astronomy Observatory, 950 N. Cherry Ave., Tucson, AZ 85719, USA}

\author{Abhijit Saha}
\affiliation{National Optical Astronomy Observatory, 950 N. Cherry Ave., Tucson, AZ 85719, USA}

\author{Shuo Yang}
\affiliation{University of Arizona, Department of Computer Science, 1040 E. 4th St., Tucson, AZ 85721, USA}

\author{Zhenge Zhao}
\affiliation{University of Arizona, Department of Computer Science, 1040 E. 4th St., Tucson, AZ 85721, USA}

\author{John Kececioglu}
\affiliation{University of Arizona, Department of Computer Science, 1040 E. 4th St., Tucson, AZ 85721, USA}

\author{Carlos Scheidegger}
\affiliation{University of Arizona, Department of Computer Science, 1040 E. 4th St., Tucson, AZ 85721, USA}

\author{Richard T. Snodgrass}
\affiliation{University of Arizona, Department of Computer Science, 1040 E. 4th St., Tucson, AZ 85721, USA}
\author{Tim Axelrod}
\affiliation{University of Arizona, Steward Observatory, 933 N. Cherry Ave., Tucson, AZ 85720, USA}

\author{Tim Jenness}
\affiliation{Department of Astronomy, Cornell University, Ithaca, NY 14853, USA}
\affiliation{LSST Project Office, 950 N. Cherry Ave., Tucson, AZ 85719, USA}

\author{Robert S. Maier}
\affiliation{University of Arizona, Department of Mathematics, 1040 E. 4th St., Tucson, AZ 85721, USA}

\author{Stephen T. Ridgway}
\affiliation{National Optical Astronomy Observatory, 950 N. Cherry Ave., Tucson, AZ 85719, USA}

\author{Robert L. Seaman}
\affiliation{University of Arizona, Lunar \& Planetary Laboratory, 1629 E University Blvd., Tucson, AZ 85721, USA}
\author{Eric Michael Evans}
\affiliation{University of Arizona, Department of Computer Science, 1040 E. 4th St., Tucson, AZ 85721, USA}

\author{Navdeep Singh}
\affiliation{University of Arizona, Department of Computer Science, 1040 E. 4th St., Tucson, AZ 85721, USA}

\author{Clark Taylor}
\affiliation{University of Arizona, Department of Computer Science, 1040 E. 4th St., Tucson, AZ 85721, USA}

\author{Jackson Toeniskoetter}
\affiliation{University of Arizona, Department of Computer Science, 1040 E. 4th St., Tucson, AZ 85721, USA}

\author{Eric Welch}
\affiliation{University of Arizona, Department of Computer Science, 1040 E. 4th St., Tucson, AZ 85721, USA}

\author{Songzhe Zhu}
\affiliation{University of Arizona, Department of Computer Science, 1040 E. 4th St., Tucson, AZ 85721, USA}

\collaboration{(The ANTARES Collaboration)}

\correspondingauthor{Gautham Narayan}
\email{gnarayan@stsci.edu}

\begin{abstract}
The unprecedented volume and rate of transient events that will be discovered by the Large Synoptic Survey Telescope (LSST) demands that the astronomical community update its followup paradigm. Alert-brokers -- automated software system to sift through, characterize, annotate and prioritize events for followup -- will be critical tools for managing alert streams in the LSST era. The Arizona-NOAO Temporal Analysis and Response to Events System (\texttt{ANTARES}) is one such broker. In this work, we develop a machine learning pipeline to characterize and classify variable and transient sources only using the available multiband optical photometry. We describe three illustrative stages of the pipeline, serving the three goals of {\em early}, {\em intermediate} and {\em retrospective} classification of alerts. The first takes the form of variable vs transient {\em categorization}, the second, a multi-class typing of the combined variable and transient dataset, and the third, a purity-driven subtyping of a transient class. While several similar algorithms have proven themselves in simulations, we validate their performance on real observations for the first time. We quantitatively evaluate our pipeline on sparse, unevenly sampled, heteroskedastic data from various existing observational campaigns, and demonstrate very competitive classification performance. We describe our progress towards adapting the pipeline developed in this work into a real-time broker working on live alert streams from time-domain surveys.
\end{abstract}


\section{Introduction}

The Large Synoptic Survey Telescope~\citep[LSST,][]{ivezic2008lsst} will revolutionize astrophysics, probing deeper than the previous generation of wide-field surveys, and replacing static maps with a continuous movie of the night sky -- and producing $\sim$20~terabytes of raw images every single night. This is approximately the same data volume as all of the imaging data obtained by the Sloan Digital Sky Survey \citep[SDSS,][]{SDSSDR14} over a decade. However, despite the dramatic increase in depth and data volume, on-going surveys including the Dark Energy Survey \citep[DES,][]{DES}, and the newly-commissioned Zwicky Transient Facility \citep[ZTF,][and references therein]{Law-2009, Rau-2009,Ofek12,ZTF} still \emph{visually} inspect candidate detections of source variability, commonly referred to as ``alerts,'' to determine the most promising targets for followup studies.

Visual inspection does have merits: humans are very capable at distinguishing pathological data from interesting astrophysical behavior, can make inferences despite sparse or missing information, and can combine and derive complex contextual information which is incorporated into their final classification decision. But as the volume of alerts grows, the efficacy of visual inspection by humans decreases, and the process of classification by visual inspection becomes increasingly inconsistent and rate-limiting. Consequently, rare and extremely scientifically interesting objects often go unstudied because detailed follow-up could not be prioritized in time, or simply because they were not identified as unusual from sparse early phase observations. 

The limitations of human inspection have been recognized for some time, but the effort to replace eyeballs with algorithms at different stages of the analysis is not a simple task. As reported by various transient surveys, candidate transient sources flagged by the difference imaging pipelines include ``bogus'' artifacts, overwhelming the number of bonafide objects detected in difference images by an order of magnitude or more. Increasingly complex automated filtering is being applied to winnow down the alert streams and separate real astrophysical sources from artifacts, e.g., with SDSS: \citealt{SDSSrealbogus}, Pan-STARRS: \citealt{PS1realbogus}, DES: \citealt{DESrealbogus}, the Intermediate Palomar Transient Factory (iPTF): \citealt{Brink-2013, Rebbapragada-2015}, Hyper Suprime-Cam Survey: \citealt{Morii-2016}, and notably Optical Gravitational Lensing Experiment (OGLE): \citealt{OGLErealbogus} using an unsupervised hierarchical self-organizing map (SOM) and the High cadence Transient Search (HiTS): \citealt{deephitscnn} using deep learning with a rotationally invariant convolutional neural network (Deep-HiTS). 

Improvements in real-bogus \emph{categorization} are necessary, but will not address the \emph{classification} of alerts that enables an investigator to select followup targets matching particular criteria. The key distinction between real-bogus categorization and alert classification is that the former functions on features extracted from individual difference images---a snapshot---whereas the latter considers the time evolution of the source---a sequence---along with contextual information. Alerts from different images, potentially from different surveys, must be cross-matched and combined, tagged with contextual information and the results of any spectroscopic followup, and this combined alert packet for each astrophysical source must be characterized, and if possible, definitively classified. Human screening of an alert package can take several seconds. LSST is expected to produce $\sim 10^7$ alerts per night; a rate that far exceeds the capacity of visual inspection. LSST will require a software system capable of 1) automated real-time classification of alerts and 2) filtering and distribution of alerts to allow astronomers to focus on objects that are relevant to their scientific interests -- an ``alert-broker.''

The Arizona-NOAO Temporal Analysis and Response to Events System (\texttt{ANTARES}) is an alert-brokering system that we are developing to meet these requirements. \texttt{ANTARES} is designed to sift through the alert stream and characterize events, with the goal of identifying phenomena that are exceedingly infrequent, as well as those of interest to the broader astronomical community~\citep{Saha14,Saha16}. This feature distinguishes \texttt{ANTARES} from existing broker services, which use human inspection to serve early-phase transient alerts to the community. While both automated classification and alert distribution systems exist, they have seldom been combined, and even the few automated alert-brokers that have been developed have never been operated at LSST scale.

\subsection{An Overview of this Paper}\label{sec:overview}
In this work, we will present progress towards developing \texttt{ANTARES}, describing some of the challenges posed by alert-broker development, as well as highlighting how a system may be useful to different studies. We illustrate this by posing three problems, focused on solving three different scientific questions:
\begin{enumerate}
  \item Real-time filtering on extremely sparse data to enable recognition, categorization, and rapid follow-up of unusual phenomena
  \item General alert characterization to provide different science interests with feeds of relevant objects
  \item Stringent retrospective classification to identify members of a specific class of object
\end{enumerate}
These three use cases drive the choices we make for feature extraction and classifier development in this work.

Following a broad review of automated classification and alert brokering in \S\ref{sec:review}, we describe the design and current status of the \texttt{ANTARES} classification pipeline in \S\ref{sec:ANTARES}. We structure the subsequent sections of this work to reflect the development cycle for our pipeline, with each section discussing a different component of the broker system. In \S\ref{sec:datasources}, we describe the datasets, real and simulated, that are used to develop and validate the classification pipeline. In \S\ref{sec:featureextraction}, we describe feature extraction from the time-series data, prior to the development of the multiple stages we use to process the light curves in \S\ref{sec:machinelearningpipe}. In \S\ref{sec:classifierperformance}, we describe the training and application of our suite of classification algorithms, and we introduce metrics to evaluate their performance. We summarize our conclusions in \S\ref{sec:conclusions} and discuss future avenues of development for \texttt{ANTARES} in \S\ref{sec:futurework}.

\clearpage 

\section{An Overview of Automated Classification \& Alert-Brokering}\label{sec:review}

We provide a brief overview of the existing literature on automated variable and transient classification systems and alert-brokering systems in the following subsections. 

\subsection{Variable Classification}\label{sec:variable-classification}
The automated classification of variable stars has a long history, beginning with the development of methods to determine the periods of pulsating and eclipsing variables, including the Lafler-Kinman statistic \citep{LaflerKinman}, the Lomb-Scargle periodigram \citep{LS1, LS2} and several Fourier power spectrum methods, Analysis of Variance \citep[AOV,][]{AOV}, Phase Dispersion Minimization \citep[PDM,][]{PDM}, Bayesian Evidence Estimation \citep{PeriodBayes}, Conditional Entropy \citep{ConditionalEntropy}, as well as hybrid methods \citep[]{SahaANdVivas}. Despite the differences in these techniques, \citet{PeriodFindingComparison} found that most methods exhibited comparable performance on realistic data, and there was no single optimum algorithm. The most accurate algorithm for period determination depended on the astrophysical class being studied -- information that is not available for most sources. Indeed, variable classification is one of the purposes for which period determination is used in the first place. 

While the period is one of the most important features in discriminating between different classes of variables, there are many features that are sensitive to light curve shape. Early work by \citet{AutoclassVariables} showed that even small sharp features in the light curves could improve the accuracy of period determination. Together with clustering techniques and naive Bayes classifiers, these light curve shape features could be used to label large datasets much faster than would have been possible by visual inspection. \citet{Debosscher-2007} successfully applied machine learning techniques to a very diverse set of variables, with over twenty different classes, drawn from several different surveys, showing that the classification features and the technique were extremely robust and could be used to label new datasets \citep{Debosscher-2009, Sarro-2009}.

\citet[][hereafter R11]{RichardsVariablesML} vastly expanded the set of features that are sensitive to the shape of the light curve to include many metrics that are more robust in the presence of noisy or spurious data. R11 found that including these robust features and a hierarchical taxonomy of labels could dramatically improve classification performance on the \citet{Debosscher-2007} dataset. \citet{Richards-2012} utilized these features, together with iterated active learning -- prioritizing followup of objects whose inclusion into the training sample maximally helps classification -- to reduce sample selection biases. R11 remains the conceptual basis for many contemporary methods applied to large datasets, such as \citet{Masci-2014}, as well as many software packages for variable star classification such as \citet[][]{Kim2015Upsilon}\footnote{\url{https://github.com/dwkim78/upsilon}} and \texttt{FATS}\footnote{\url{https://github.com/isadoranun/FATS}} and its derivative, \texttt{feets}\footnote{\url{https://github.com/carpyncho/feets}}. Unfortunately, as many of the observational programs that discover variable stars only observe with a single filter, many of these software packages are designed with the assumption of single-band photometry. These packages do not make use of multi-color information, despite its utility in discriminating between different classes of variables.

\subsection{Transient Classification}\label{sec:transient-clasification}
While variable classification is retrospective -- it can be applied well after the observations have been taken -- one of the main goals of transient classification is to operate in \emph{real-time} to select objects for spectroscopic follow-up \emph{while they are still active}. Early ``flash'' spectroscopy is particularly important for understanding the physics of the progenitor systems \citep[e.g.,][]{khazov2016}. Even programs that derive cosmological constraints from photometric samples of SNIa use spectroscopy to assess their contamination \citep{Jones-PS1Phot}, and have a relatively narrow window to obtain spectroscopy with sufficient $S/N$. 

This need for real-time rapid response drove some of the earliest advances in using machine-learning techniques for transient classification. \citet{Poznanski2007} attempted to distinguish between SNIa and core-collapse SNe using single-epoch photometry along with a photometric redshift estimate from the probable host-galaxy. Many contemporary techniques such as \citet{onlygalclass} and the \texttt{sherlock} package\footnote{\url{https://github.com/thespacedoctor/sherlock}}, can operate on sparse, or only contextual information, allowing for spectroscopic follow-up while the transient rises to maximum light. 

The Supernova Photometric Classification Challenge (hereafter SNPhotCC) in 2009 was one of the earliest efforts to employ machine learning algorithms for light curve classification along the lines of the work done on variable stars. The challenge simulated supernova light curves with the properties of the DES, and aimed to determine which techniques could distinguish SNIa from several other classes of supernovae. The data for the SNPhotCC was provided in full along with the original types chosen for generation by \citet{Kessler2010Announcement}.

The techniques used to classify for the challenge varied widely, from basic spline fitting and semi-supervised learning, to much more complex methodologies that fit light curves with a variety of templates (such as \citealt{guy10}, or parametric descriptions such as \citealt{Karpenka-2013}), and compared classification results using an ensemble of classification schemes \citep{Kessler2010Results}. A measure of the value of this exercise is that the SNPhotCC dataset is still used as the reference standard to benchmark contemporary supernova light curve classification schemes, such as \citet[][hereafter L16]{Lochner2016}. In this paper, we integrate the non-parametric transient classification framework developed in L16 as a component of the \texttt{ANTARES} pipeline. An updated version of SNPhotCC, now including more classes than just different types of supernovae, and with simulations appropriate for LSST, the ``Photometric LSST Astronomical Time-series Classification Challenge'' (PLAsTiCC), is in development\footnote{\url{https://plasticcblog.wordpress.com/}}.

\subsection{Alert Filtering and Distribution Systems -- Brokers}
An alert-broker is a software system to rapidly characterize and filter alerts. Brokers must be capable of producing large and pure samples of known astrophysical classes, and must therefore be effective at distinguishing between classes, as well as being able to identify rare and novel sources within each class. As they operate in real-time, alert-brokers must cope with complex \emph{streaming input} from different astronomical facilities studying different parts of the electromagnetic spectrum. This can include objects with pathological and/or missing data and that can be contradictory - e.g. spectra of the same object from two groups with two different classifications. The data for each object will span a different range in phase, so unlike classification algorithms, assembling a consistent input feature vector across all events is simply not possible. This means brokering systems must necessarily adopt some sort of hierarchical classification scheme, depending on the amount of data available for a given object. This reflects the R11 finding that adopting a hierarchical taxonomy of variable stars improves classification.

There are a few notable examples of alert-brokers that have classified variables and transients in real-time on survey data. The \texttt{Oarical} system \citep{BloomRichards12} employed during the PTF survey was one of the first automated transient broker. However, alerts were not made public, but rather supplied to PTF members through ``Marshals,'' -- alert brokering systems tailored to the needs of specific projects within the PTF collaboration. These alerts were still visually inspected before follow-up. The \texttt{Oarical} code has also never been made public, and while elements of it have been adapted to iPTF alerts and will likely be applied to ZTF alerts, the Marshals still employ visual classification to select targets for spectroscopic follow-up. The Catalina Real-time Transient Survey (CRTS) operated an automated classifier \citep{CRTSbroker} for a period in 2015; however, current optical transients are all reported as human classified. \texttt{4PiSky}\citep{4piSky}\footnote{\url{https://4pisky.org/}} and the (now defunct) SkyAlert \citep{SkyAlert} serve alerts in the VOEvent format\footnote{\url{http://www.ivoa.net/documents/VOEvent/index.html}} \citep{voevent} defined by the International Virtual Observatory Alliance.

Brokers do not require machine learning classification schemes at all, and indeed many brokering systems simply annotate the alert stream. One of the most successful, though underappreciated, brokering systems is the Rochester Supernova web page\footnote{\url{http://www.rochesterastronomy.org/supernova.html}} maintained solely by amateur astronomer David Bishop. The Rochester page has provided alerts on potential supernovae to a vast number of follow-up teams for more than two decades, and is often updated with the results of follow-up observations before the International Astronomical Union's (IAU) Transient Name Server (TNS)\footnote{\url{https://wis-tns.weizmann.ac.il/}}, itself another example of a transient alert-broker. However, the critical issue for large survey projects like LSST in the coming years is having a way to cope with the quantity and the rate of data. Brokers must be able to utilize the capacity they have to keep pace with the data rate from LSST, which will produce a new image every 37 seconds. To contend with the high event rate and volume, we will have no alternative but to adapt to using machine learning algorithms to process the alert stream.  

\section{\texttt{ANTARES}}\label{sec:ANTARES}

\texttt{ANTARES} consists of two components: 1) a pipeline that will use automated classification techniques to provide real-time characterization and annotation of alerts and 2) an alert distribution system that will allow these annotated alerts to be searched and filtered by astronomers.

The classification pipeline must be capable of handling data that is incomplete, most often because of the limited observing cadence and lost observing time due to poor weather conditions. It must be able to make a preliminary classification using only the beginning of the light curve, in order to enable rapid spectroscopic follow-up and find interesting new classes of transients, as well as to probe unexplored regimes of the evolution of known classes of objects. Even without high-confidence photometric classification, criteria such as the association of contextual information, and categorization can all be used to select objects for early-time studies.

The brokering system wraps around the classification pipeline. It ingests alerts into the system, and controls the flow of data through the pipeline. The brokering system also coordinates the book-keeping of the system: the storage of extracted features and annotations in a database, as well as serving the annotated alert stream to downstream brokers, publishing rare and interesting alerts to web pages, and allowing users to search and filter the database of all annotated alerts. 

Building this system requires expertise from both astronomers and computer scientists. We began active development of the project in December 2014, adapting existing platforms and services wherever possible, and constructing new tools where none were available. The current architecture of \texttt{ANTARES} is depicted in Fig.~\ref{fig:architecture} and described below.

\subsection{Architecture of \texttt{ANTARES}}\label{sec:archreview}

\begin{figure*}[htpb]
  \centering
  \vspace{-1.0em}
  \includegraphics[height=\textheight]{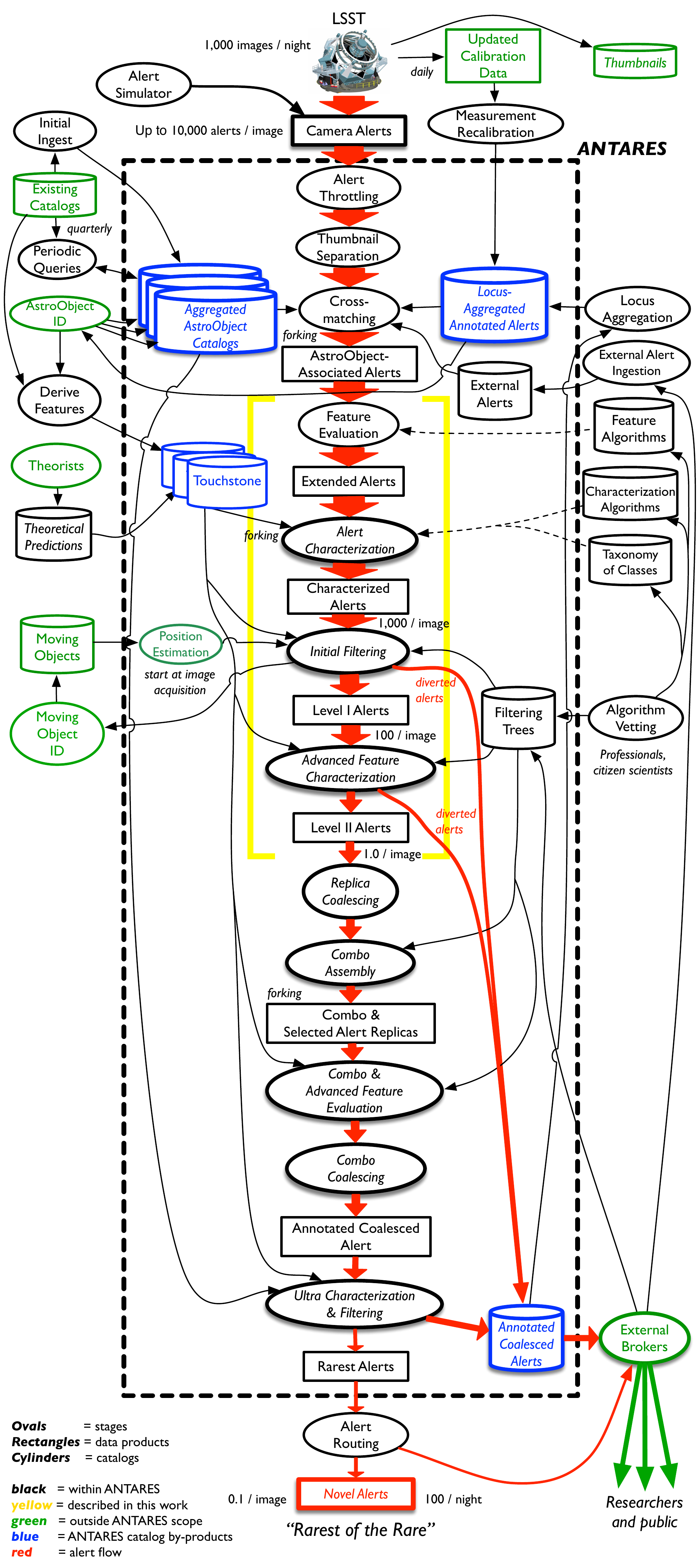}
  \caption{Schematic of the \texttt{ANTARES} architecture. The processing pipeline is enclosed by the black dashed line. The core machine learning stages described in this work and depicted in Fig.~\ref{fig:pipeline} are bracketed in yellow.}\label{fig:architecture}
\end{figure*}

The \texttt{ANTARES} system consists of several components (any shape with a black outline in Fig.~\ref{fig:architecture}) to process alerts from LSST. Some of these components are part of the pipeline (encompassed by the dashed black outline) run on every alert package from LSST. Components that are interfaces to the community, or external systems and are not run on every alert package are outside this outline. 

Prior to ingestion by \texttt{ANTARES}, alert packets from LSST will be filtered through the LSST alert real-bogus discrimination system and the LSST moving object pipeline, to remove difference imaging artifacts and sources that exhibit significant motion against the background sky. The contents of an LSST alert packet are described in the LSST Data Products Definition Document\footnote{\url{http://ls.st/dpdd/}}. The filtered difference image alerts produced by the LSST pipeline are ingested into the \texttt{ANTARES} pipeline (the dashed outline) at the top. The alert stream can be throttled by $S/N$ if the volume exceeds our processing capacity. Extraneous information, such as difference imaging thumbnails are separated from the alert packet. Alerts cross-matched against static catalogs (referred to as ``AstroObjects''), as well as a database of previous history at the same position on the sky (a ``locus''), including any prior calculations and evaluation by \texttt{ANTARES} itself. Features are derived from all the alerts. The contextual information and computed features are stored together with the original data in the LSST alert packet as an ``extended alert.'' Where a feature or contextual attribute cannot be uniquely determined, a duplicate of the alert is created (indicated in Fig.~\ref{fig:architecture} by the text-label ``forking''), and these different copies are assigned the different possible values of the feature or contextual attributes. We term these different copies ``replicas.'' 

All the replicas of all the alerts are passed on to the various filtering and characterization ``stages.'' Some of these stages score and label the replicas, while other stages trigger actions if the replicas meet certain predefined conditions of interest. These stages (denoted by the yellow bracketed region in Fig.~\ref{fig:architecture}) are largely designed to compare the alert to a library of astrophysical events with similar properties and known classifications -- the ``Touchstone.'' Some stages are more computationally intensive than others and it would be prohibitive to run these stages on all the alerts from LSST. However, these stages are designed to do more fine-grained classification (for e.g., determining the type of supernova, rather than discriminate between supernovae and variable stars), and only need to be run on a subset of the alerts that match certain criteria after initial filtering. We ``divert'' alerts that do not meet these criteria, and do not execute these computationally intensive stages on them. We term the alerts that are retained for processing after the initial diversion ``Level I Alerts,'' and the alerts that are retained after the computationally intensive ``Advanced Feature Characterization'' stages are executed, ``Level II Alerts.'' After all the stages are run, all of the replicas of each alert are coalesced, and the ensemble of the classifiers is used to annotate the alert, thereby accounting for different possibilities even when attributes or contextual information cannot be uniquely determined. It is these core classification stages that are the focus of this work. 

In some cases, alerts may prove interesting in combination with other external alerts at the same location on the sky. For example, a tidal disruption event (TDE) may be indicated by an optical trigger in a galaxy that has not previously exhibited AGN activity, but has strong ongoing soft X-ray or IR emission. We term such structures ``combos'' and define much more specialized filtering stages to process them as needed. These stages can be more computationally intensive as the volume of alerts being processed decreases with each filtering stage of the pipeline. Expected numbers of alerts per image are indicated by the text near components in Fig.~\ref{fig:architecture}, as well as the decreasing width of the red arrows between stages. Most alerts will not meet the criteria necessary to trigger the creation of a combo, and will simply pass through these stages.

The distribution system of \texttt{ANTARES} consists of stages that deliver our processed alerts to the astronomical community (these stages are depicted having a red arrow crossing the dashed black outline). Alerts that are labelled are diverted and recorded along with any computed features into the database, while the subset of the that appear different from all known classes represented in the Touchstone library are identified as novel and prioritized for rapid follow up. All annotations for all alerts are stored in the database and are accessible by end users (illustrated in Fig.~\ref{fig:architecture} by the blue catalog labelled ``Annotated Coalesced Alerts'').

Various elements of the analysis are not provided directly by \texttt{ANTARES}, and the system instead interfaces with LSST or other facilities that provide these features (indicated in green in Fig.~\ref{fig:architecture}). These include image calibration and subtraction products, updates to catalogs, lists of moving objects, etc. We plan to develop an API that will allow users to ``daisy-chain'' instances of \texttt{ANTARES}, refining the output from the general broker to produce results specific to their scientific interests. We will create a mechanism for users to further process and interact with the annotated alerts and features stored in our database via a Project Jupyter Hub, to serve many use cases where real-time access to the alert stream isn't required.

\subsection{Current Status}

Many components of \texttt{ANTARES} have been developed: an alert simulator to inject simulated data into the system, relational databases to store external catalogs, ingested data, and processing results, an API for the execution of the different stages and to interact with the database, a load-balancing system for parallel execution, web frameworks to inspect the results, systems for configuration management and tracking provenance, as well as the front-end interfaces to serve this data to the community. Development of these components required only a relatively small amount of astrophysical data to serve as test cases. At present, we are moving from our initial goal of identifying the rarest of the rare (where we focus on the lowest red box in Fig.~\ref{fig:architecture}) to the much bigger challenge of developing a general purpose broker. In this paper, we have elected to focus on the core classification stages, as this remains one of the largest open research questions involved in the development of an alert-broker. 

\section{Data Sources for Developing and Testing \texttt{ANTARES}}\label{sec:datasources}

In order to develop and test the machine learning-based classification stages of the \texttt{ANTARES} broker (the yellow bracketed region in Fig.~\ref{fig:architecture}) and provide a framework for future classification using LSST's data products, we need much larger datasets that provide a representative range of astrophysical sources. To capture the diversity of the variable and transient sky, we drew from three separate data sources described below.

\subsection{OGLE Variable Stars}
The Optical Gravitational Lensing Experiment \citep[OGLE,][]{OGLE} is a wide-field sky survey originally designed to search for microlensing events. The project monitors over 200 million stars over several years, assembling an enormous database of photometric measurements. The OGLE project classified their sources using several techniques, evolving from human inspection, through simple categorization and template fitting, to machine learning-based techniques over their more than two decades of operation. The OGLE-III Catalog of Variable Stars \citep{OGLE3CVS} consists of the observations in $V$ and $I$ collected beginning in 2001. We augment the OGLE-III release with new objects from OGLE-IV, available through their FTP server\footnote{\url{ftp://ftp.astrouw.edu.pl/ogle/ogle4/}}, using a custom parser to translate the files into a standard format, and with a standard set of labels. Adding the new objects allows us a larger sample from classes that are underrepresented in the OGLE-III release. 

\begin{figure*}[htpb]
  \centering
  \includegraphics[width=\textwidth]{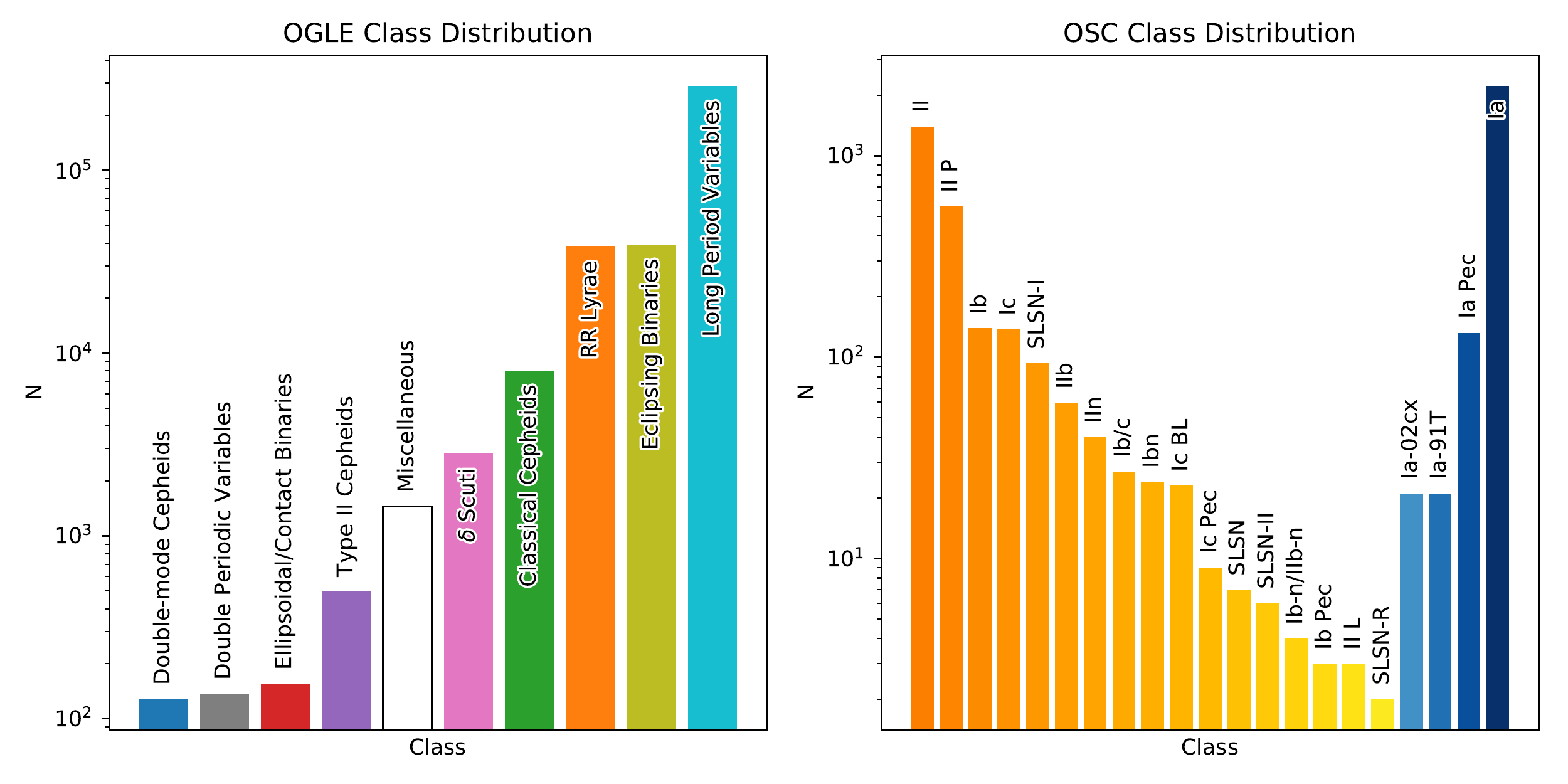}
  \caption{Class distributions for the OGLE (left) and OSC (right) datasets. These distributions clearly highlight the imbalance in the datasets.}
  \label{fig:classdistrib}
\end{figure*}

The classes (with catalog designations listed parenthetically) represented in the OGLE sample include classical Cepheids (``cep''), double periodic variables (``dpv''), $\delta$~Cepheids (``dcep''), type II Cepheids (``t2cep''), ellipsoidal/contact binaries (``ell''), eclipsing binaries (``ecl''), $\delta$~Scuti variables (``dsct''), RR Lyrae of different types (aggregated as ``rrlyr''), long period variables or Miras (``lpv''). Miscellaneous other types are aggregated together (``misc'') and are not used in this work, often because there are insufficient observations with $S/N > 5$ to derive features reliably. There is a significant class imbalance in this dataset, with over two orders of magnitude more long period variables than type II Cepheids or double periodic variables. The distribution of class labels for the OGLE is shown in the left panel of Fig.~\ref{fig:classdistrib}. 

Any representative smaller sample drawn from the OGLE data that is sufficiently large to train machine-learning algorithms is also inevitably much larger than the entirety of the labelled supernova dataset because of the difference in rates. Despite these caveats, OGLE is by far the largest multi-class dataset of labelled variable stars with photometry in two passbands (i.e. have at least some color information). 

\subsection{The Open Supernova Catalog}
The second dataset considered is drawn from the repositories of the Open Supernova Catalog \citep[OSC,][]{Guillochon17}. The OSC is a public online repository that accepts observations of supernovae from all willing contributors. This public repository operates with the goal of containing a complete collection of publicly available data of supernovae with both spectra and photometry in visible and near-visible wavelengths, as well as radio and X-ray. However, the dataset is heterogeneous, comprising different classes of supernovae, from a variety of teams, selected with different strategies, observed using different sites, telescopes, instruments, passbands, cadences and image reduction pipelines, and subject to weather losses, photometric calibration errors, and mislabelling. This heterogeneity is why it has never been used by classifiers until this work. The data stored are stored in the repository as they were originally reported, and include outliers and missing data. Even processing all these varied light curves to produce a consistent feature vector for each object is a non-trivial challenge.

We used all OSC light curves available on the repository as of January 2017. We sub-selected objects from 1987 to the present that met a few quality control criteria to exclude objects with an insufficient number of observations. We only used light curves that had at least one claimed type as at least one label is necessary for all supervised machine learning algorithms. We required at minimum 25 observations across all bands to ensure there is at least some color and phase information. For objects where the type is disputed, we take the type claimed by the largest number of unique sources, and if this is insufficient, we assert the most recently claimed type is correct. The resulting data draws from many different references and a disparate collection of surveys.  

In many cases, particularly with the OSC, many of the class labels are sub-types of a parent class, while others are simply ill defined. For example, ``Ia-Pec'' in the OSC sample includes under luminous and over luminous SNIa while ``Ia-02cx'' and ``Ia-91T'' are taken to be under luminous and rapid declining, and over luminous respectively, and there is no precise criteria for putting an object in one class vs the other. Additionally, in the OSC, many classes are represented by 10 or fewer members -- wholly inadequate for training any classifier. We aggregate the many sub-types of supernovae into two broad classes -- Ia and non-Ia (see \S\ref{sec:sne_classif}). The distribution of class labels for the OSC dataset is shown in the right panel of Fig.~\ref{fig:classdistrib}.

\subsection{SNPhotCC Simulated DES light curves}
While the real light curves of OSC provide a true assessment of classifier performance, the simulated SNPhotCC dataset is the reference standard for supernova classifier performance assessment (see \S\ref{sec:transient-clasification}). It consists of approximately 18,000 $griz$ light curves of types Ia, Ib/c, and II supernovae that were simulated to match the expected properties of DES. Each type is represented in accordance with its expected volumetric rate. For the purposes of the challenge, a ``spectroscopically confirmed'' subset was provided to give a training set for classification. The training sample is simulated to match the properties of dataset with photometric observations on a 4 meter telescope with a limiting $r$-band magnitude of 21.5, and spectroscopic followup with an 8 meter telescope with a limiting $i$-band magnitude of 23.5~\citep{Kessler2010Announcement}. This roughly models the performance of the on-going DES project, based on a library of historical conditions at the Cerro Tololo site assembled during the Equation of State: SupErNovae trace Cosmic Expansion (ESSENCE) survey \citep{Narayan16}.

The simulated Type Ia SNe are generated using empirically derived models. Many of the non-type Ia supernovae were provided by (in increasing order of redshift) the Carnegie Supernova Project~\citep[CSP,][]{Folatelli10,Stritzinger11}, the Sloan Digital Sky Survey~\citep[SDSS II,][]{Holtzman08}, and the Supernova Legacy Survey~\citep[SNLS,][]{sullivan11,betoule13}, and at the time of the challenge, were unpublished. The set of non-Ia light curves undersamples the potential variety that will be seen in future large scale surveys, with under 50 objects providing the base model for all simulated non-Ia events. However, the SNPhotCC dataset still provides a useful step in the interim period before more complete data become available. The Pan-STARRS Medium Deep Survey (MDS) and DES has taken the requisite observations to provide such a sample, and will likely be included in the PLAsTiCC dataset. 

The SNPhotCC provided two separate challenges, one with host-galaxy redshifts (``+HOSTZ'') and one without (``-NOHOSTZ''). Unsurprisingly, all of the methods in \citet{Kessler2010Results} performed better with host-galaxy redshifts. In this work, we have chosen to work on the set \emph{without} host-galaxy redshifts, to reflect that the southern sky that LSST will scan has considerably shallower galaxy catalogs than the northern sky, which has already been imaged by SDSS and PS1. 

\subsection{Differences between the Datasets and Pre-Processing}

\begin{figure*}[ht]
 \centering
 \includegraphics[width=0.9\textwidth]{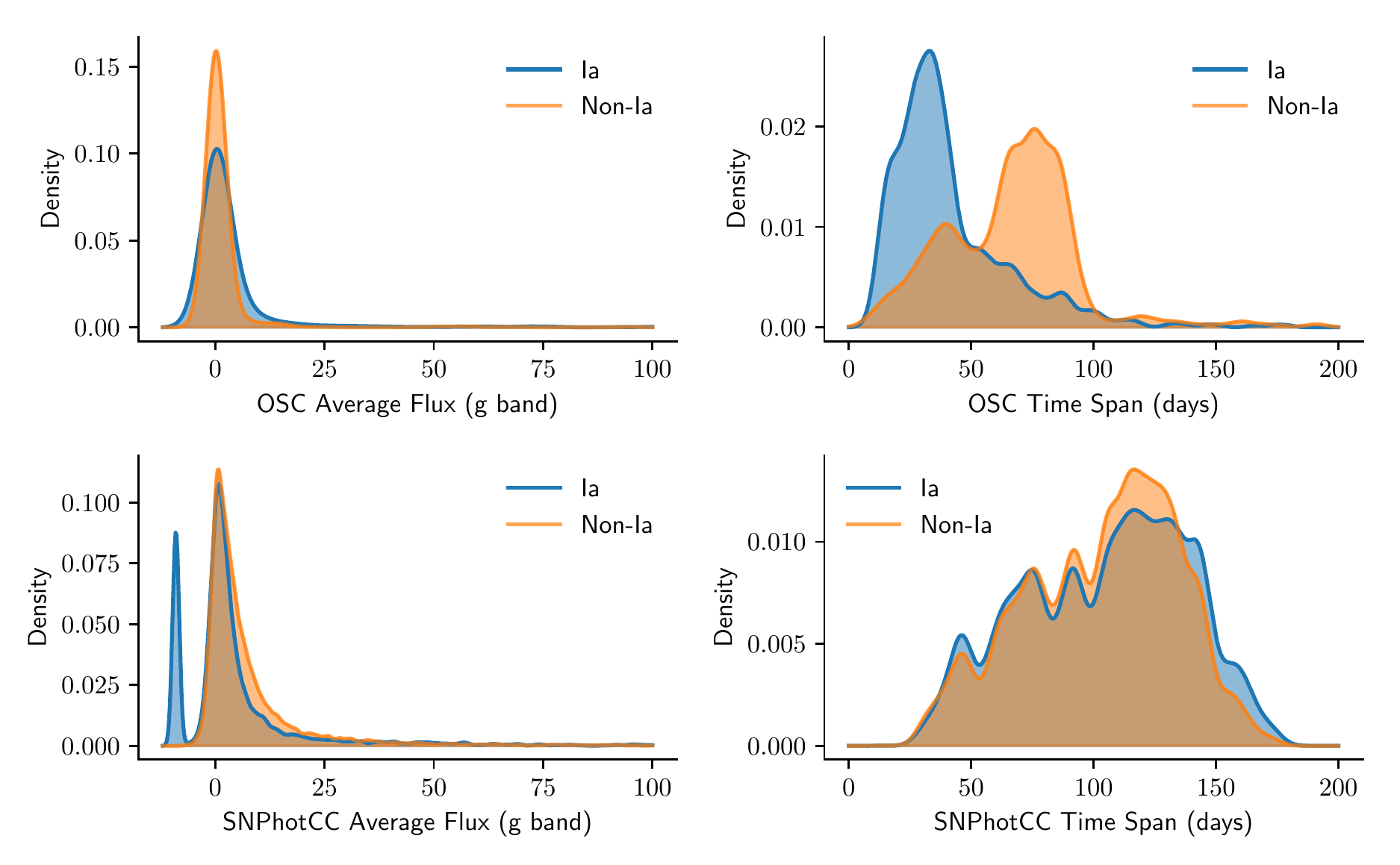}
 \caption{Density plot of the average magnitude ({\it left}) and time span ({\it right}) of light curves from the OSC ({\it upper row}) and SNPhotCC ({\it lower row}) datasets.}
 \label{fig:density_OSC_DES}
\end{figure*}

The density plots of the average magnitudes and time spans of light curves shown in Figure \ref{fig:density_OSC_DES} highlight key differences between these data sources. The SNPhotCC exhibits no strong differences between the Ia and non-Ia for both features. Overall their time spans are longer than the OSC supernovae, as is to be expected from simulating an untargeted survey, following a source for as long as it is visible. For the OSC, though the average magnitudes are similar for Ia and non-Ia light curves, the time spans differ greatly, reflecting the different follow-up strategies used by surveys for these different classes. We explore differences in classifier performance between the OSC and SNPhotCC dataset further in \S\ref{sec:classifierperformance}. LSST will produce a much larger and homogeneous collection of supernova light curves to replace these disparate datasets. By assessing performance on existing surveys, we can set a lower bound on the classification performance, as well as create a tool that may be used by on-going precursor surveys to LSST, such as CRTS, the All-Sky Automated Survey for Supernovae (ASAS-SN), and the ZTF. 

Most data need relatively simple pre-processing. We remove observations with negative or zero observed flux uncertainty, observations which have missing values for the flux or flux uncertainty (including those represented with dummy values such as -99), and observations represented only by upper detection limit or lower flux limit, rather than a measurement of the flux.

\subsubsection{Passband Mapping}\label{sec:bandmapping}

In order to allow for objects from different surveys in the OSC to be comparable to one another, a reference set of filters must be defined. For consistency with the SNPhotCC data, the $griz$ system used by the DES telescope was adopted as the photometric system. All passbands in the OSC were mapped into the $griz$ system using a simple heuristic: if the passband of the original survey overlapped significantly with any of the DES filters, then we assigned the observations to that band. We developed an assessment of the quality of the photometry of various low-redshift supernova surveys, based on their cadence, and their typical phase coverage. In cases where multiple filters from different low-redshift surveys mapped to the same standard band, we adopted the photometry from the higher ranked survey. The surveys are listed in ranked order in Table~\ref{tab:surveys}. In order to combine the OSC data with the data from the OGLE survey, which has observations only in $V$ and $I$, we use just the OSC data mapped into the $g$ and $i$-band for variable-transient categorization and classification. The OGLE variables are diverted after this stage. We use all available $griz$ information in the final stage of our pipeline, where we are compare supernovae against each other to determine the feasibility of selecting a pure sample of SNIa.

\begin{table}[ht]
 \centering
 \begin{tabular}{p{1.3in}|c}
 \toprule
 Survey/Telescope      & Largest Compilation \\ \tableline
 CfA Supernova Group     & \citet{Hicken12} \\
 Carnegie SN Project   & \citet{Stritzinger11} \\
 Lick Observatory SN Search (LOSS) & \citet{Ganeshalingam10} \\
 NASA Swift Telescope   & \citet{Brown09} \\
 CfA-IR & \citet{Friedman15}       \\
 SDSS SN Search & \citet{Holtzman08}   \\ \tableline
 \end{tabular}
 \caption{The surveys represented in the OSC light curves that require passband mapping (see \S\ref{sec:bandmapping}). The surveys are in decreasing order of precedence for resolution of conflicts when multiple observer-frame passbands from different surveys can be mapped to the same DES passband for any given object. Here, the order roughly reflects the $S/N$ and the cadence of the observations.}\label{tab:surveys}
\end{table}

This is a very simplistic system for passband mapping, and it throws away a large number of high quality light curves in the process when there is no appropriate overlapping passband, as well as losing information from each object. There are more sophisticated techniques available for passband mapping such as \citet{ScolnicSupercal}, but these require information about the redshift and/or type of the object. The former is not typically available, and the latter is precisely the quantity that we wish to infer. While the SNPhotCC provides reference data with a consistent photometric system, more than 80\% of the light curves (both type Ia and non-type Ia) are synthetically generated using the same empirically derived models that are often used later in classification. Thus, it remains necessary to use datasets such as the OSC, even though the disparate nature of the sources makes comparative analysis more challenging. We intend this exercise to be illustrative of broker development, as well as a step towards semi-supervised learning on hitherto unclassified PS1 and iPTF light curves, and is therefore necessary, despite these compromises. 

\section{Feature Extraction Methodology}\label{sec:featureextraction}

Supervised learning algorithms are trained on a library of features from objects with known class labels to derive a desired function. For classification purposes, this function is the class label itself. The derived function can then be applied to the same features derived from data with unknown class labels to predict the value of the function.

Supervised learning algorithms therefore require three successive operations: 
\begin{enumerate}
 \item Projecting the high dimensional information of each source to a lower dimensional feature space that encapsulates as much information as possible: \textit{encoding}
 \item Learning a metric that can be used to quantify the differences between classes in that space using many labelled instances of each class: \textit{training}
 \item Applying the metric to new unlabelled instances in order to filter, characterize or classify them: \textit{prediction}
\end{enumerate}

In this section, we detail the first of these operations -- the encoding, or feature extraction. The choices of the features that we attempt to extract are driven by the information available for each source. For this work, we consider feature extraction in three distinct regimes, corresponding to the three different questions we posed in \S\ref{sec:overview}:
\begin{enumerate}
 \item Soon after the alert is issued, where only limited information is available -- perhaps only a couple of observations. The only features derived at this stage are an amplitude, rate of change, or a single color. Rapid real-time prioritization will depend on effective filtering in this regime, and will rely on additional contextual information.
 \item At intermediate times, when a few tens of observations exist across all bands. This amount of data is sufficient to derive descriptive statistics, such as the kurtosis and skewness, and to attempt to characterize significant timescales, but the sources continue to evolve in time, and the observations have not covered the entire phase curve. Characterization and broad classification in this regime can still support target selection for follow-up observations, but not rapid early-time studies.
 \item At later times, when the observations across all bands span the entire phase curve, and there is perhaps little significant additional information to be gained from further observations. Classification in this regime is effectively retrospective. However, prompt publication by the broker system may still have important advantages for many science objectives.
\end{enumerate} 
\clearpage
\subsection{The Variability Probability Distribution Function: Thresholding the Variability of the Galactic Stellar Background}\label{sec:vpdf}

When an alert is initially issued by LSST, it includes previous photometry of the source, if any, in the preceding 100 days. It is expected that forced photometry for the same time period at the location of the source will be available only within 24 hours. For many transient sources, this photometry will only consist of detection limits. Nevertheless, despite this sparse data, alert-brokers must be able to determine if the source's variability is significant, and worth additional follow-up. The presence of an associated galaxy near the location of the source will greatly aid in this determination, but a galaxy is not always guaranteed to be present (e.g., the source might be at high redshift, and a low surface brightness galaxy may not be detectable, or a deep catalog of static sources may not be available at the alert location), or may not be uniquely identified. 

To identify and prioritize interesting transients at early phases, a broker must be able to distinguish them against the \emph{background of stellar variability}. As the time-series of the parts of the sky scanned by LSST builds up, the bulk of the variable sources for the first few years of LSST operations will be Galactic stars, AGNs and asteroids without known orbital elements that have been misclassified as stationary difference image objects \citep[][hereafter SR14]{Ridgway-2014}. Such objects may be of interest to several groups. However, in any search for rare transient events, these relatively mundane sources can be considered contaminants. The alert rates for \emph{new sources} will plummet as the survey progresses, while the a
alert rates of new, early-phase transients will remain constant (or improve, as various improvements are made to the survey). We can maximize the early science return from LSST, and reduce the number of alerts to process, and thereby the load on brokers, with effective early characterization and filtering. This, in turn, requires a description of the background variability.

\begin{figure}
 \centering
 \includegraphics[width=0.47\textwidth]{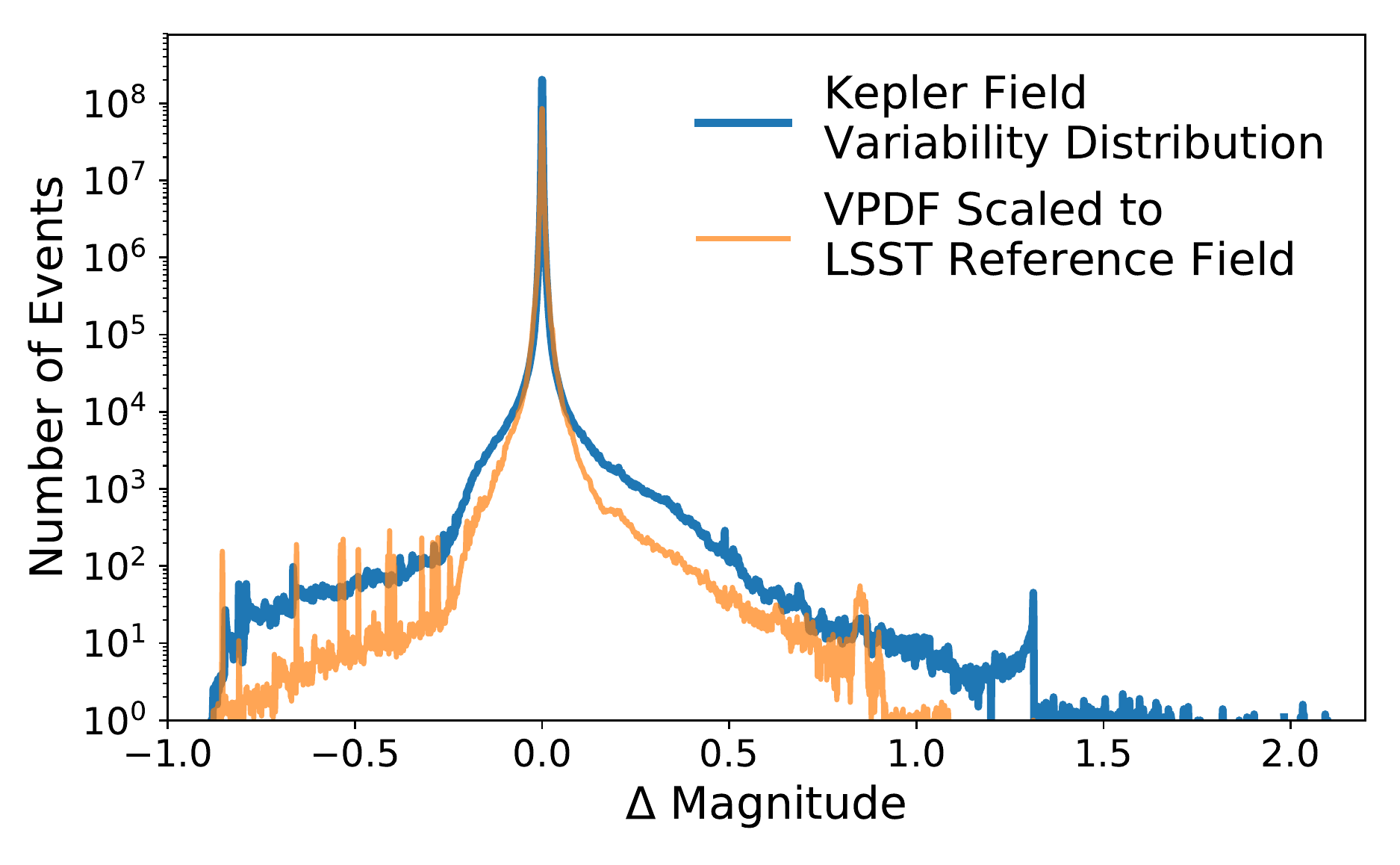}
 \caption{Distribution of variability of stars in the Kepler field (blue) obtained from the Quarter 6 time-series data of more than 155,000 stars as a function of change in magnitude from the median. The orange curve shows the resulting distribution after scaling to the Galactic stellar population at the LSST reference field centered on R.A.$=0\degree$, Dec.$=0\degree$, based on simulated stellar catalogs from the Besan\c{c}on Galaxy model.}
 \label{fig:VPDF}
\end{figure}

SR14 define the Galactic variability probability distribution function (VPDF) using time-series data from \mbox{\textit{Kepler}} Quarter 6, of over 155,000 stars belonging to different spectral types. The distribution gives the probability of seeing a given variability expressed as root-mean-square (RMS) brightness amplitude for the particular stellar type. However, the \textit{Kepler} field of view covers only a small region on the sky, and the underlying stellar distribution varies with position in our Galaxy. For a given LSST pointing/image covering about $10~{\rm deg}^2$ along a particular line of sight, a distribution of the spectral types of the Galactic stellar population is simulated using the Besan\c{c}on model\footnote{\url{http://model.obs-besancon.fr/}} \citep{Robin-2003}. The variability distribution from the Kepler field is then scaled to the corresponding stellar population of the pointing, an instance of which is shown in Fig.~\ref{fig:VPDF}. There are limitations with using the \textit{Kepler} sample, as it does not include supergiants or white dwarf stars, and there are only a few examples of giant stars. However, unlike other studies that follow stars known to exhibit variable behavior (such as OGLE), the \textit{Kepler} sample contains light curves from a large distribution of stars with \emph{unbiased} selection, and is therefore the best currently available dataset for this analysis. The final data release from the \textit{GAIA} mission \citep{GAIA} will include all epoch photometric catalogues, and will hence supersede the \textit{Kepler} sample. 

Matheson et al. (2018, in preparation) use the \emph{change in magnitude}, rather than the RMS brightness amplitude, to define the VPDF. This parametrization is \emph{always} useful for a source, even with a single significant observation, as the source required some change in brightness, i.e., a non-zero amplitude in the difference image analysis, to have been triggered as an alert. It can be viewed as a Bayesian prior, formalizing the use of other features that have been used to identify significant variability with sparse data, such as the median absolute deviation (MAD). Rather than a simple threshold for all sources, the VPDF gives the likelihood that the variability of a source at some location on the sky is significantly different from the expected background Galactic stellar variability in the same region. This likelihood can be reduced to a simple binary label (significant/not significant variability).

A variation of the above filtering algorithm is to consider the probability distribution of changes in magnitude, $\Delta m$, for a given window between the observations, $\Delta t$. This expands the parameter space along the time axis, and has been explored for real-time classification of alerts by some groups (for example \citealt{Mahabal-2011}). Using light curves of different classes of variable stars and transients obtained with the CRTS survey, they constructed the joint-probability distribution of $(\Delta{m}, \Delta{t})$ for each class and used a machine learning-based algorithm to classify a new alert with an assembled time-series. More recently, \citet{Mahabal-2017} created a new implementation based on these two parameters, mapping each light curve into the $\Delta m$-$\Delta t$ space, which in turn was used to construct a 2D feature vector. Such $\Delta m$-$\Delta t$ ``images'' of light curves belonging to different classes of variable stars were then used to train a convolutional neural network, often used for image recognition and classification in astronomy \citep[e.g.,][]{Dieleman-2015, Jacobs-2017}, and achieved acceptable results for some classes. 

We are also developing an algorithm based on features extracted from the $\Delta m$-$\Delta t$ density distributions of the various classes of varying sources (Soraisam et al., 2018, in preparation), to be implemented as a stage within \texttt{ANTARES} to help in timely characterization of interesting/novel events. For example, in this work, among the different kinds of variable stars and transients considered, supernovae serve as an example of the cream of the crop. Most of the low-redshift supernova light curves in the OSC are not from untargeted searches, but rather they are observed after being discovered by targeted searches. Consequently, the first observation is on average only a week before maximum light, after the source has risen significantly, and even with a conservative estimate for the previous non-detections, these light curves would trivially show high significant variation relative to the Galactic VPDF. 

During regular operations we will use filter stages, such as VPDF thresholding within \texttt{ANTARES}, to flag alerts that exhibit significant variability compared to the expected Galactic stellar variability background. These filters use contextual information immediately, without having to train on observations, allowing them to be applied to new surveys that are just coming online. This level of filtering also gives us a throttle; we can raise and lower the threshold dynamically, depending on the region of sky under consideration, or if the alert volume for an image becomes very high, e.g., as a result of poor image subtraction, with many unfiltered artifacts.

Another generalization would be to add the ability to threshold, not just on the background stellar variability, but also based on the source's own previous history. If \texttt{ANTARES} has classified a source, we could in principle skip processing of new alerts at the same location until a sufficiently large number of new observations are obtained to merit repeating feature extraction and source characterization. Such an approach implicitly assumes that the current alerts for a source do not add significant new information over the existing history of LSST observations; each source has an ``envelope of mundanity'' describing an expected range of feature space for new alerts based on previous observations. However, we could generalize VPDF to trigger on alerts that significantly exceed not just the background stellar variability, but also this envelope. Such a stage could detect sources that suddenly begin behaving atypically, \'a la KIC 846285 or Tabby's star. This approach would allow a quantitative treatment of what is presently a subjective assessment that astronomers make for sources that are acting ``weird.''

\subsection{Timescale Characterization}\label{sec:timescalefeature}

As the source evolves, and more observations become available, characterization can become increasingly sophisticated. The bulk of the work of a broker will be with objects in this intermediate regime, as classification here serves the needs of much of the astronomical community including cone searches and queries against catalogs (e.g., after a gravitational wave trigger is issued), querying for objects similar or different from some source (e.g., to make a comparison plot), monitoring a previously discovered transient for abnormal behavior (e.g. a supernova re-brightening such as iPTF14hls, \citealt[][]{iPTF14hls}), building target lists for follow-up studies (e.g., a project to determine the metallicity of RR Lyrae in the Galactic Halo), among many other use cases. 

Where the VPDF and $\Delta{m}$-$\Delta{t}$ formalism compare objects based on the timescales probed by the survey, the object itself may have several characteristic timescales. We use the Lomb-Scargle algorithm \citep{LS1, LS2} to determine a characteristic timescale for the observations in each passband of each source. While these timescales should agree, and be equal to the fundamental period for periodic sources, the periods determined from different bands may differ due to aliasing. In the case of transients, there is no a priori reason for the timescales computed in different bands to agree. 

\begin{figure*}[htpb]
 \centering
 \includegraphics[width=\textwidth]{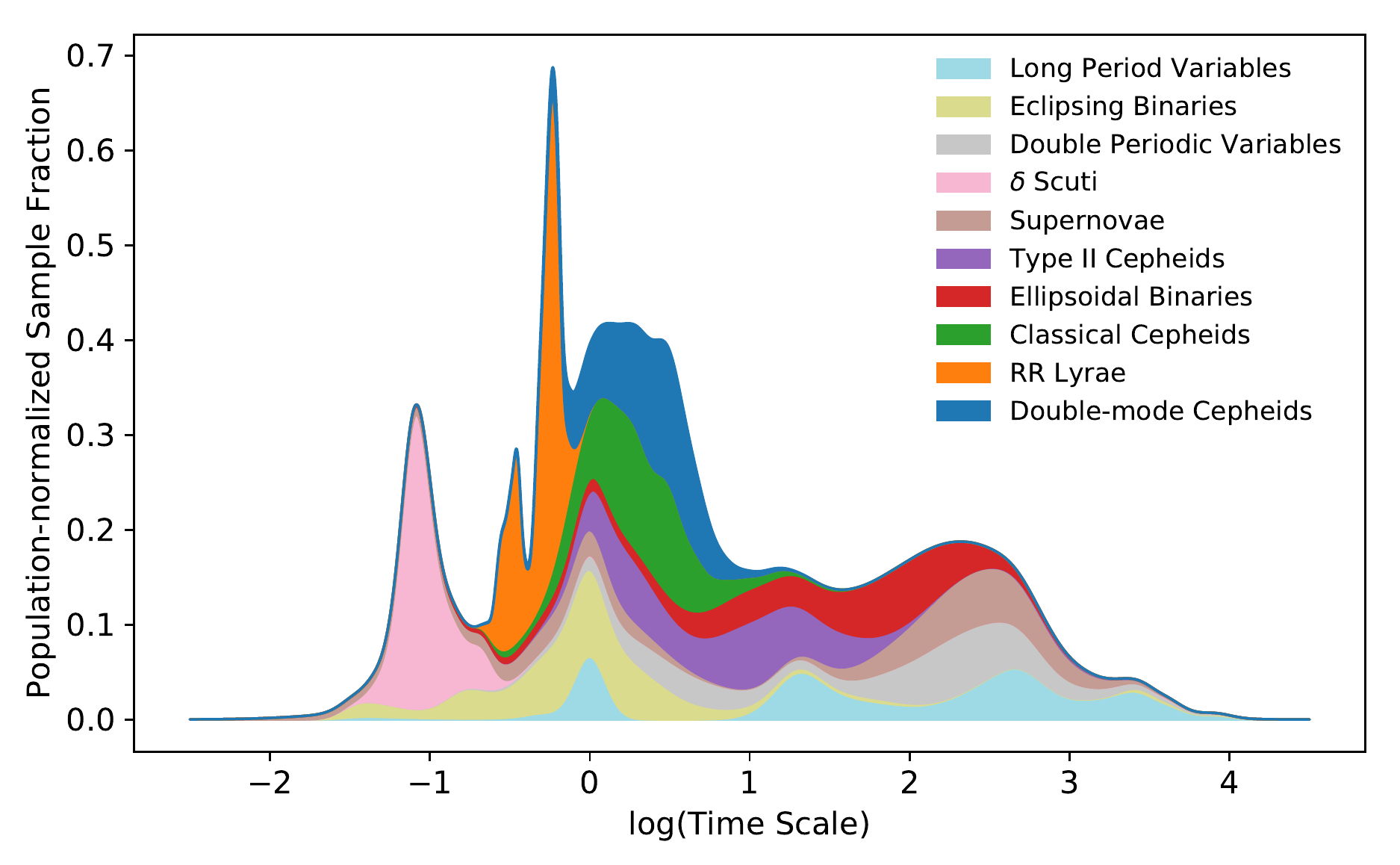}
 \caption{Logarithm of the characteristic timescales (cf.~\S\ref{sec:timescalefeature}) in days for different classes of objects, plotted as stacked kernel density estimates -- a population-normalized cumulative distribution.}
 \label{fig:timescale}
\end{figure*}

We use the inverse of the frequency with the maximum normalized Lomb power as the characteristic timescale in each band. We compute the average of the entire power spectrum, as well as the standard deviation, and determine the frequency around the peak where the normalized power drops below one standard deviation above the average. We use the half-width of this range as the uncertainty in the timescale. This is also a useful feature in classification, as the typical cadence will always probe a few cycles for variables with short periods, and these will have a relatively narrow peak in the Fourier power spectrum, \emph{even if the correct period is not determined but an alias is selected from the periodogram}. By contrast, rising transients are akin to windowed impulse functions, and have a broad spread, and therefore much larger uncertainties. Additionally, we define the period $S/N$ ratio as the difference of the peak power and the median power, normalized by the standard deviation of the power spectrum, and compute the logarithm of the False Alarm Probability \citep[FAP,][]{Baluev08}. These four quantities are computed for all the OGLE objects in both $V$ and $I$, as well as the OSC objects in all bands. In both cases, we only use the data that can be mapped to $g$ and $i$, as described in \S\ref{sec:bandmapping}.

We use these features in our machine learning pipeline (\S\ref{sec:machinelearningpipe}) in two distinct stages that operate in sequence. In the first, we use all eight timescale characterization features (four each for the $g$ and the $i$ band) to make an initial variable-supernova binary classification. In the second stage, we adopt the features of the band with the smaller time scale uncertainty (in this work, between $g$ and $i$ as those are the only bands available in our combined dataset) as the time scale feature vector for the object. We combine these time scale features with more descriptive statistics to train a multi-class variable-supernova classifier, which in addition to labeling different variable classes, can be used to identify any SNe wrongly classified as variables in the first stage. This structuring enforces a hierarchical taxonomy of classes, with early classification errors detected and fixed by later stages as more data become available. 

Gaussian kernel density estimates of the distribution of time scales that were adopted for the second stage for different classes are shown in Fig.~\ref{fig:timescale}. As can be seen in the figure, intrinsic periodic variables show sharp peaks with low dispersion in their distribution, whereas extrinsic variables and supernovae have many different time scales in the sample. The distribution of the shaded regions gives a sense for relative fractions within each class that exhibit different characteristic time scales, while the offsets between shaded areas between any two classes indicate how easy it is to distinguish them from each other. For example, a simple cut on the timescale at 1 day, would be sufficient to distinguish $\delta$ Scuti variables from supernovae of all types, but it would be difficult to use period alone to distinguish between different types of Cepehid variables. Note that even in this one dimensional feature space, it is possible to see clear subgroups, e.g., among the RR Lyrae, corresponding to RRab and RRc, even though the label aggregates both subtypes.

A more general technique for timescale characterization would be to simply use the discrete Fourier transform power spectrum evaluated over a fixed range of frequencies as the feature vector. However, with the heterogeneous data we are utilizing in this work, the power spectrum serves primarily to distinguish different surveys with different observing cadences that were targeting different classes of objects from each other, providing excellent accuracy but without affording insight into the physical differences between classes. 

We use the \texttt{fasper} routine implemented by the \texttt{FFTW}\footnote{\url{http://www.fftw.org/}} library \citep{fftwcode} to derive the periodogram, which we wrap in \texttt{ANTARES} with the \texttt{pyFFTW}\footnote{\url{http://hgomersall.github.io/pyFFTW/}} module. We've found that we can derive a more robust estimate of characteristic timescales for periodic variables in our dataset using the power spectrum computed using the \emph{multiband} Lomb-Scargle algorithm implemented in the \texttt{gatspy}\footnote{\url{http://www.astroml.org/gatspy/index.html}} package \citep{gatspycode}. However this improvement is due to the package's implementation of the regularization scheme described in \citet{gatspy}, which is appropriate for variable sources but may not be appropriate for transient sources. Consequently, we only derive one characteristic time scale per band in this work. 

\subsection{Magnitude Distribution Characterization}\label{sec:magcharacterization}

As the number of observations grow, we can define more features that can help in classification by incorporating information on the light curve and the flux distribution. The goal of feature extraction is to describe the light curves of the OGLE and OSC sources, but without introducing features that allow a distinction to be made on apparent magnitude (which would introduce bias) or on the observational properties of the survey, such as the telescope and instrument (which would not be informative). Many of these features have already been used very successfully for the classification of variable stars (see references in \S\ref{sec:variable-classification}). We elected to adopt many of the features in R11 as well as \citet{SIDRA} and \citet{Kim2015Upsilon}; these are listed in Table~\ref{tab:featuredescription}.

\begin{table*}[ht!]
 \centering
 \begin{tabular}{l|c}
 \toprule
 Feature & Description \\
 \tableline
  Autocorrelation Integral ``$\alpha$'' & The integral of the correlation vs time difference following \citet{SIDRA} \\ 
  Entropy & The Shannon entropy assuming a Gaussian CDF following \citet{SIDRA} \\
  HL Ratio & The ratio of the amplitudes of points higher and lower than the mean \\
  Inter Quartile Range (IQR) & The difference between the $75^{\text{th}}$ and $25^{\text{th}}$ percentile of the magnitude distribution \\
  Kurtosis & Characteristic ``peakedness'' of the magnitude distribution \\
  Median Absolute Deviation (MAD) & A robust estimator of the standard deviation of the distribution \\
  Shapiro-Wilk Statistic ``$w$'' & A measure of the magnitude distribution's normality \\ 
  Standard Deviation/Mean ``$\sigma_{m}/\langle m \rangle$'' & A measure of the average inverse $S/N$ \\
  Skewness & Characteristic asymmetry of the magnitude distribution \\
  Stetson K & An uncertainty weighted estimate of the kurtosis following \citet{Stetson-96} \\
  Von-Neumann Ratio ``$\eta$'' & A measure of the autocorrelation of the magnitude distribution \\
 \tableline
 \end{tabular}
 \caption{Description of the 11 features extracted from observations of each passband of all sources. Additionally, four passband-independent time scale features are computed (see \S\ref{sec:timescalefeature}), producing an $N_{\text{pb}} \times 11$ + 4 dimensional feature vector for every object, where $N_{\text{pb}}$ is the number of distinct passbands.}
 \label{tab:featuredescription}
\end{table*}

\begin{figure}[ht]
  \centering
  \includegraphics[width=0.5\textwidth]{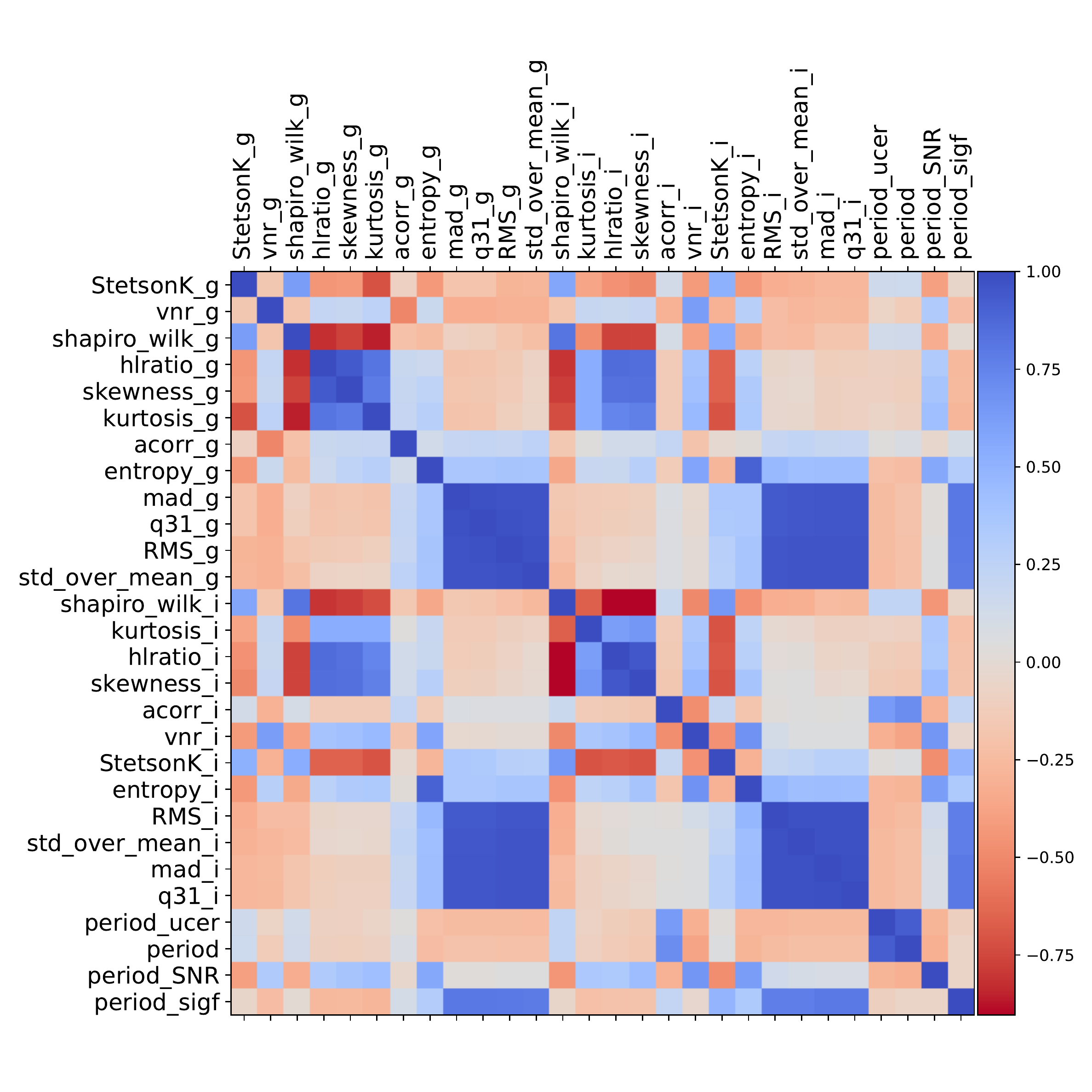}
  \caption{Correlation matrix of the dataset. Features that characterize the magnitude distribution extracted from the two bands in this work, $g$ and $i$ have many features that are strongly correlated within and between each band, with the colorbar at right displaying the coefficient of correlation.}
  \label{fig:PCA}
\end{figure}

The elements of the feature vector are generally correlated (see Fig.~\ref{fig:PCA}), while many machine learning algorithms are designed with the expectation that each element is an independent variable. While feature extraction is intended to lower the dimensionality of the signal, it is possible to reduce the dimensionality even further. We expect correlations, both within the feature vector of each band (as many of the features are effectively different measures of the same quantity), and between bands (as the behavior of astrophysical sources is correlated across bands). 

We scale the entire feature matrix by removing the mean and de-correlating or ``whitening'' \citep{whitening} the data -- scaling each column to be uncorrelated and have unit variance. We then use a standard principal component analysis (PCA) to reduce the dimensionality of the feature matrix. We use a $N$=15 dimensional vector of the PCA features. This choice explains $\approx96$~\% of the sample variance. \\ \\ 

\subsubsection{Using t-SNE as a Feature Space Visualization Tool}\label{sec:class_imbalance}

\begin{figure*}[ht]
 \centering
 \hspace{-20pt}
 \includegraphics[width=\textwidth]{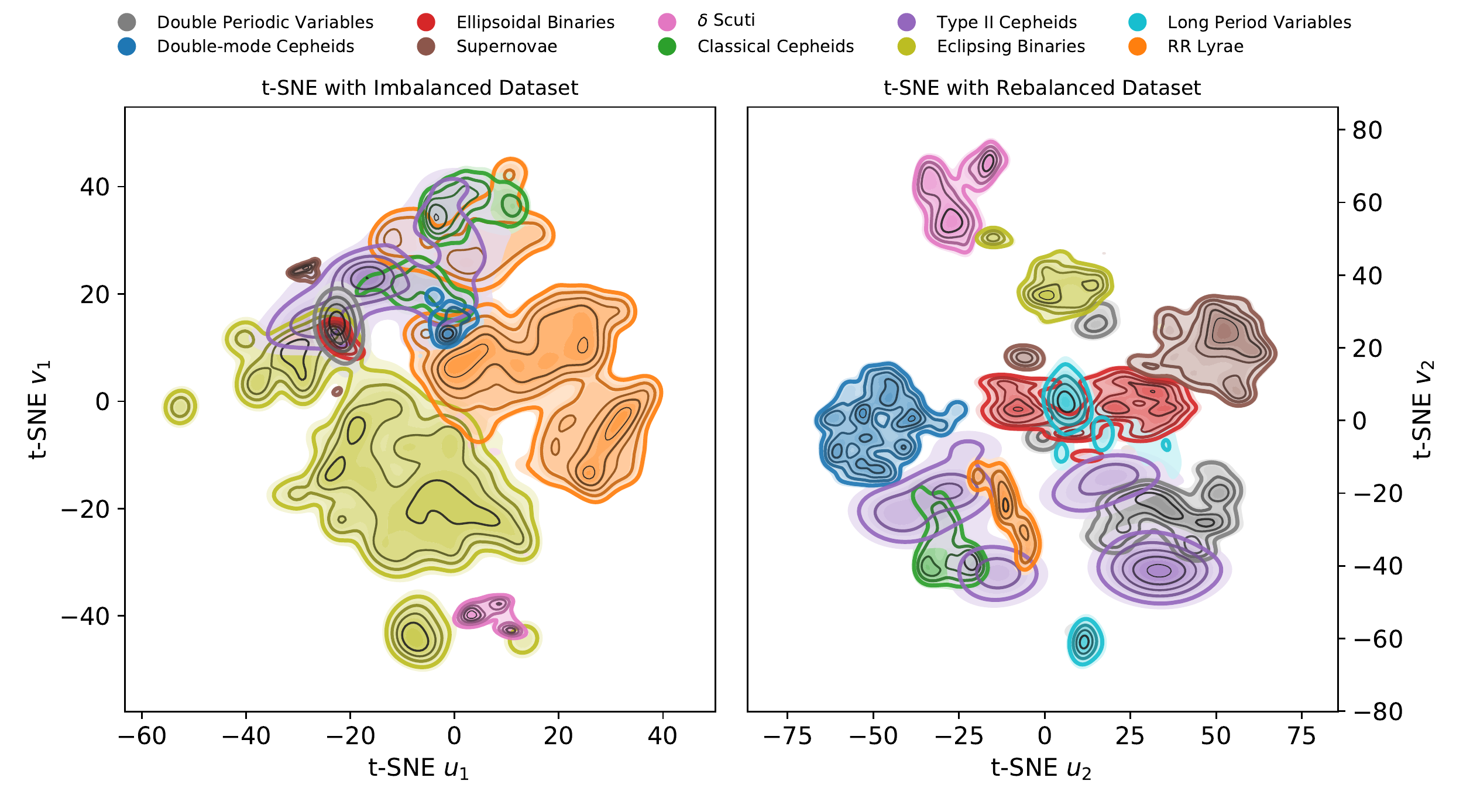}
 \caption{A t-distributed Stochastic Neighbor Embedding (t-SNE) ``Petri dish'' of the feature matrix after scaling and PCA with whitening on the imbalanced (Left) and rebalanced (Right) dataset. The t-SNE embedding attempts to preserve points that are similar to each other in the full feature space, clustering them together in a lower dimensional space that is easier to visualize. The t-SNE embedding is determined without using any class labels, and points are colored by class after they are embedded in the low dimensional space. We represent the $\sim 10^{2}$--$10^{5}$ points in each class using a bi-variate kernel density estimate (KDE). The two t-SNE axes are not physically interpretable, nor are the sizes of the clusters. With the class imbalance, the t-SNE has performed relatively poorly, dispersing the largest classes (eclipsing binaries and RR Lyrae) and separating them from each other, but without distinguishing between the minority classes. We account for the drastic class imbalance in the real datasets using a combination of techniques in \S\ref{sec:machinelearningpipe}. When applied to a balanced dataset, the t-SNE separation improves dramatically, with most of the classes clearly distinguished.}
 \label{fig:t-SNE}
\end{figure*}

While the explained variance from the principal component analysis is a measure of how well the low-dimensional feature vectors represent the higher dimensional input, it does not help us determine whether the features themselves are predictive -- if they can be used effectively for classification. That question will ultimately be answered by training and validating the machine learning classifiers, however, even without machine learning, we can examine if the feature vectors are likely to be useful by constructing a t-Distributed Stochastic Neighbor Embedding \citep[t-SNE,][]{maaten2008visualizing}\footnote{\url{https://lvdmaaten.github.io/tsne/}}.

We use the \texttt{multicore-tsne} implementation\footnote{\url{https://github.com/DmitryUlyanov/Multicore-TSNE}} \citep{multicore-tsne} of the Barnes-Hut variant of t-SNE algorithm \citep{BHTSNE}, which can produce a 2- or 3-D representation of a high-dimensional space, clustering similar points together. The algorithm constructs a k-D tree of all the points, and computes the Euclidean distance between each point and its $k$ nearest neighbors using a Student-t distribution to convert this distance into a probability that the two points are similar. 

The algorithm then attempts to find a low-dimensional space that preserves this probability, using a gradient descent algorithm, minimizing the sum of the Kullback-Leibler divergence (a measure of the divergence between two distributions). The number of nearest neighbors is an input to the algorithm known as ``perplexity'', however it is largely insensitive to this choice, and we obtain very similar embeddings for perplexity between 100--300, such as that in Fig.~\ref{fig:t-SNE}. The algorithm is unsupervised - i.e it constructs clusters of similar points \emph{without any knowledge of the labels}. Classes that are well separated in a t-SNE visualization can generally be distinguished from each other by a machine learning classifier, however the converse does not always hold. 

There are several caveats to t-SNE visualizations, and neither distances between nor the sizes of the clusters may be informative \citep{wattenberg2016how}\footnote{\url{https://distill.pub/2016/misread-tsne/}}. Additionally, the stochastic nature of the algorithm means that different runs with the same data, or different partitions of the input set with different class balance can produce different, although qualitatively similar results. We therefore use the algorithm for visualization rather than classification.

We use our PCA feature vector (described in \S\ref{sec:magcharacterization}) as the input to the t-SNE as these data are largely insensitive to survey characteristics and the quality of the light curves. Using the light curves directly would require that all objects be interpolated on to a common grid, and the resulting t-SNE would likely be much more sensitive to gaps in the light curve or differences in $S/N$, which could lead to clustering that does not reflect astrophysical differences between classes. We examined several embeddings derived from our dataset, and found that this clustering is not perfect, and some groups such as RR Lyrae appear to be divided into subgroups or clumps. We visually inspected members of the subgroups, and find this division to be a reflection of reality, with the t-SNE separating RRab from RRc and RRd subtypes, even with the relative low dimension of the input feature set. Additionally, while it was not possible to distinguish between the different subtypes of Cepheid variables (classical, double-mode and type II) using only their characteristic time scale, as seen in Fig.~\ref{fig:timescale}, these groups are distinguishable in the t-SNE plot. 

The t-SNE also throws into sharp focus the scale of the imbalanced learning problem. It was necessary to suppress long period variables in the visualization, as they outnumber the next largest class by an order of magnitude, and the structure was dominated by clusters of Miras. By default, many machine learning algorithms will optimize ``accuracy'' -- the overall fraction of the predicted labels are correct -- at the expense of ``recall'' or ``sensitivity'' to smaller classes. Simply, in a very imbalanced dataset with e.g., 997 members of a red class, and only 3 members of a black class, classifiers are very likely to simply ``put it all on red'', even if the most interesting events are the rare ones. This has implications for several problems, from fraud detection in financial data, to identifying electromagnetic counterparts for gravitational wave sources. 

The t-SNE is best employed as a diagnostic tool to identify potential classification challenges with the dataset and the chosen feature representation, prior to machine learning. This can help avoid three common cascaded problems often encountered in applied machine-learning work: the adoption of naive classification algorithms, the use of inappropriate metrics, and the fine tuning of the algorithms to optimize those metrics. 

\subsection{Light Curve Characterization}\label{sec:advancedcharacterize}
Once sufficient observations have been obtained to cover the source's phase curve fully, complex feature extraction can be performed, even if computationally intensive, as the object is unlikely to require any reprocessing. Classification of transients in this regime is retrospective, as they will have faded to below the level of the sky. Nevertheless, classification at these times is extremely important and can serve a variety of research programs. In this section, we consider a project with one such requirement -- the need for an extremely high purity sample of SNIa for cosmological studies, extracted by determining the subclass for the supernova output of the previous stages, e.g., by the LSST Dark Energy Survey Collaboration (DESC). For this level of classification, we follow the approach of L16 and construct a much more complex feature vector to describe the events, using Gaussian process regression followed by wavelet decomposition to model the events.

Observations from real astrophysical surveys are unevenly spaced, have gaps due to weather losses, and have heteroskedastic errors with faint objects having much lower $S/N$ than bright objects. Even with perfect data, objects in the distant Universe are redshifted and undergo time-dilation, making a direct comparison with low-redshift objects non-trivial. A precise comparison requires knowledge of the ``K-correction''~\citep[originally described in][]{oke68}, but this cannot be computed without knowledge of the underlying spectral energy distribution (SED) of the object, and implicitly, its astrophysical class, which is the same quantity that we wish to infer from the data. Furthermore, many feature extraction and machine learning algorithms impose additional requirements on the data, such as evenly spaced observations, and the absence of missing data.

To make use of these algorithms, we use various methods to generate smooth, evenly sampled representations from the noisy, sparse observations. Because alert-brokers such as \texttt{ANTARES} will not have a priori information about the class of the object under consideration, these methods cannot rely on templates of various astrophysical classes, or on simple parametric representations of light curves, as none are sufficiently general to describe all possible variable and transient phenomena. Additionally, these methods must be robust enough to work despite the limitations of observations, and must be computationally efficient to work at scale with the LSST data rate. 

\begin{figure*}[htpb]
 \centering
 \includegraphics[width=\textwidth]{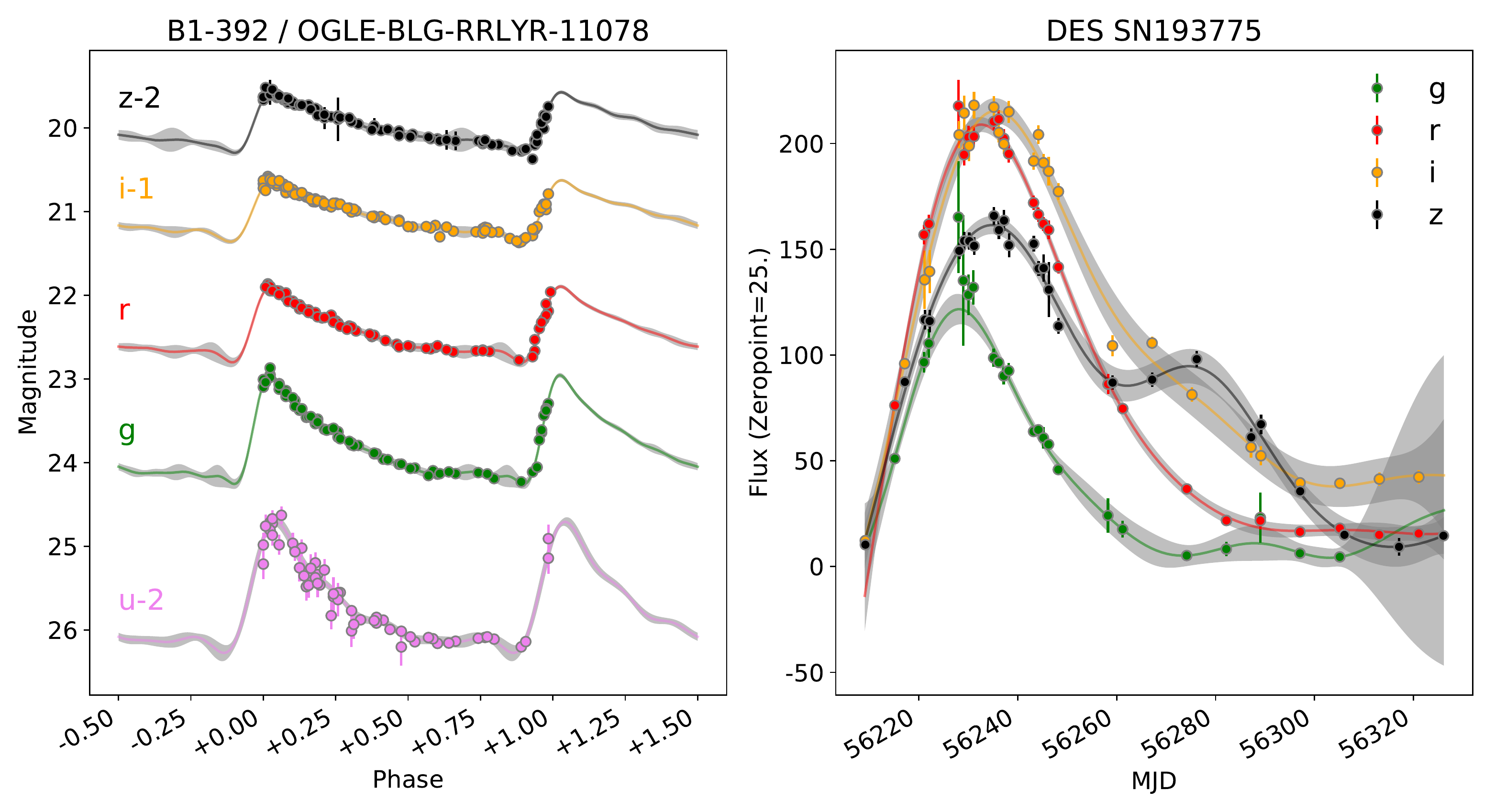}
 \caption{Gaussian processes are flexible enough to produce a smooth representation of multiband light curves of both periodic and transient objects reported in flux or in magnitudes using appropriate kernels. This is illustrated with model fits for an RR Lyrae (left) from a survey of the Galactic Bulge with DECam (courtesy of P.I., A. Saha) and a simulated SNPhotCC supernova (right). The RR Lyrae, B1-392, is chosen as it appears in \citet{SahaANdVivas}, and is also present in the OGLE dataset (OGLE-BLG-RRLYR-11078) used in this work, while the SNPhotCC light curve has a redshift of 0.4089, near the median redshift of the DES and Pan-STARRS SNIa samples.}
 \label{fig:GP_mod}
\end{figure*}

\subsubsection{Gaussian Process Regression}\label{sec:gpmodeling}

A Gaussian process \citep{GPs} models every point in some continuous input space (time in the case of light curves) with a normally distributed random variable, that defines the distribution for how far the point may lie from the mean. Any finite collection of Gaussian random variables follows a multivariate normal distribution. Conditioning this model on the observations -- the regression -- solves for the mean of this multivariate normal distribution as a function of the continuous input variable. 

Additionally, the regression problem is simplified by imposing a parametric function relating two points in the input space, $t_i$ and $t_j$, to each other -- a covariance model. Often, this ``kernel'' function is expressed solely in term of the separation of the two points, $|t_i - t_j|$; in this case, the covariance model is described as ``stationary''. This parametrization can be quite flexible as any linear combination of kernels remain a valid kernel. Consequently, the Gaussian process framework is very adaptable and is used to describe a wide array of data from different fields. We used the \texttt{george}\footnote{\url{http://george.readthedocs.io/en/latest/}} python module \citep{Ambikasaran2014} to perform the Gaussian process regression for each supernova light curve, adopting the smooth Matern~$3/2$ kernel. The output is insensitive to the choice of the kernel parametrization, with Matern~$5/2$ and squared exponential kernels performing very comparably, and the differences between the outputs are comparable with the uncertainties on the data. 

While Gaussian processes are a very general technique, and can be used for sophisticated modeling of observations while accounting of the measurement uncertainties, we use it as a generalized method of interpolating the light curve observations on to a common grid. Gaussian processes are computationally intensive ($\mathcal{O}(N^3)$ for a set of $N$ observations), the limited number of observations in a typical light curve and simplifications we can make by modeling 1-D time series data with stationary kernels serve to keep the computational cost extremely small. Example Gaussian process regression models are shown in Fig.~\ref{fig:GP_mod} for an SNPhotCC Ia as well as a periodic variable star.

\subsubsection{Wavelet Decomposition}\label{sec:wavelet_transforms}

\begin{figure*}[htbp]
 \centering
  \subfloat{\includegraphics[width=0.49\textwidth]{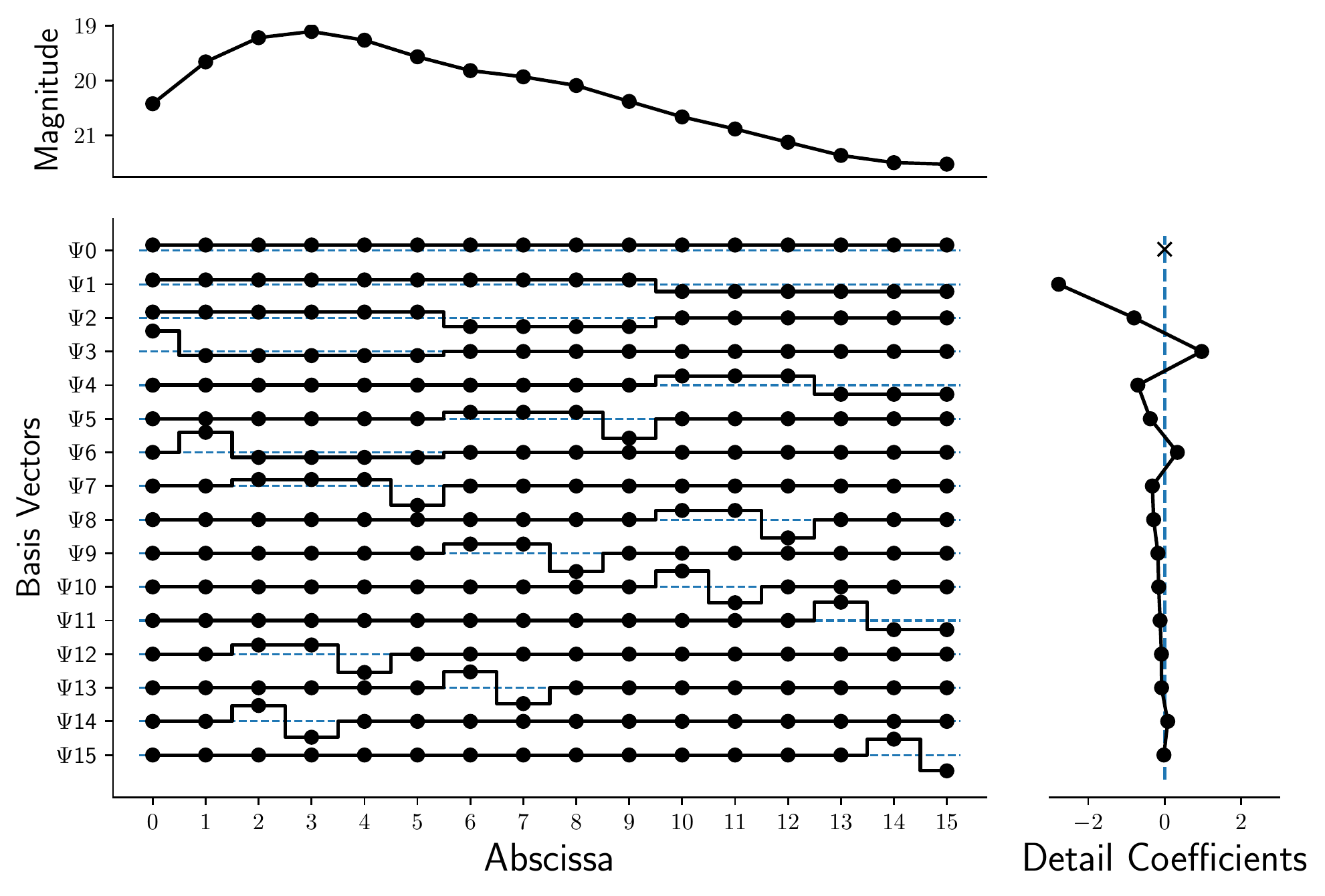}}
  \hfill
  \subfloat{\includegraphics[width=0.49\textwidth]{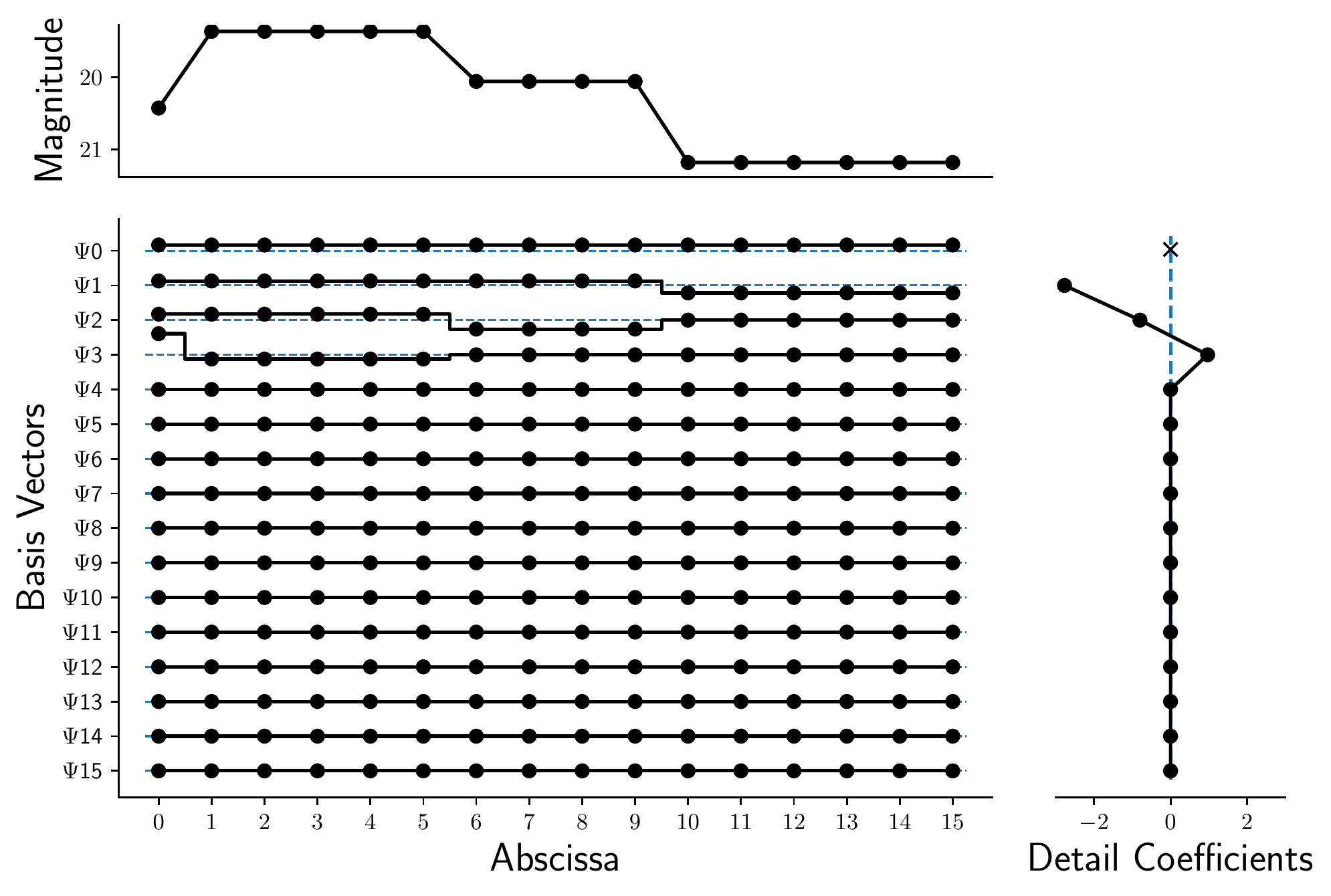}}
  \caption{The BAGIDIS decomposition (Left) and reconstruction (Right) of a Type Ia light curve. The unbalanced Haar basis vectors are shown on the left side of each plot, and the breakpoints correspond to where a given basis vector changes in sign. The corresponding detail coefficients are shown on the right. The reconstructed BAGIDIS curve is a coarse approximation, even taking only the first 4 coefficients and setting the rest to zero. Note that the first component captures the average value of the light curve, and is not used for classification of supernovae as this would introduce a cosmological bias.}
  \label{fig:bagidis_reconstruction}
\end{figure*}

Wavelet transformations are a common technique used to express a square integrable function as an orthonormal series of wavelets -- functions that begin at 0, increase, and then fall back to zero. The technique can be considered as a harmonic analysis, with complex signals expressed as the sum of a series of simple pulses. The transformation preserves the overall \emph{shape}, but not the time extension. This allows signals with the same shape but different characteristic timescales to be compared to each other using only the coefficients of the orthonormal series (frequently called the ``detail coefficients''). There are several different families of wavelets to construct the orthonormal series, each of which has different properties. The choice of which family of wavelets to use is typically made to allow an approximate reconstruction of the original signal with only a few terms of the full series, i.e., with only a few detail coefficients. Consequently wavelet transformations are frequently employed for lossy compression of the signal. 

Wavelet algorithms have been tested extensively on the SNPhotCC dataset by \citet{Varughese2015} and L16. This is the first work to validate wavelet-based methods on real observations. We use the \texttt{PyWavelets}\footnote{\url{https://pywavelets.readthedocs.io/en/latest/}} package within our pipeline, and we explore two wavelet decomposition techniques for selecting the resulting feature set to be used in the machine learning classifier: 1) the discrete wavelet transformation with BAGIDIS used in \citet{Varughese2015} and 2) the stationary wavelet transformation used in L16. 

The BAGIDIS (Basis Giving Distances) methodology \citep{TimmermanVonSachs2010} uses a basis pursuit algorithm based on the methods of \citet{Fryzlewicz2007}. The basis pursuit method decomposes the smooth input light curve onto an unbalanced Haar wavelet basis. The associated basis coefficients are ordered according to the strength of their effect on the shape of the signal. This method allows a small number of initial coefficients to encode the bulk of the information, as illustrated in Fig.~\ref{fig:bagidis_reconstruction}.

The second wavelet decomposition method the Stationary Wavelet Transform (SWT), widely used for edge detection and noise-reduction of images. For a stationary wavelet transformation $x$, with $x_n$ coefficients, a new filter $x^{m}_{n}$ is obtained by inserting $2^{(m - 1)}$ zeros or ``holes'' between each $x^{\text{th}}$ coefficient, effectively applying a sequence of low-pass filters operating on different scales to the input. The advantage of the SWT method is that it can be used with different families of wavelets and possesses translational invariance (the input abscissa can be shifted by a common offset). We elected to extract wavelet features using only the Daubechies and symlet wavelet families. Symlets were used in L16, and only differ slightly (less asymmetric) from the more common Daubechies family. The presence of hundreds of correlated wavelet features would simply introduce variance into any machine learning classifier. Consequently, we use PCA for dimensionality reduction of the SWT coefficient feature space as described in \S\ref{sec:hyperopt}. \\ 

\section{Machine Learning Pipeline}\label{sec:machinelearningpipe}

\subsection{Stages: The Functional Unit of Alert-Broker Pipelines}

As described in \S\ref{sec:ANTARES}, the \texttt{ANTARES} pipeline as a whole is comprised of a handful of discrete stages, each of which specifies a set of actions the broker must perform on the sources in the alert stream. Our architecture (see Fig.~\ref{fig:architecture}) defines different sequential levels of processing. The stages within each level can be run in parallel, with the output being coalesced, allowing each alert to be annotated by multiple processes, and providing each subsequent level with more information. Sources that do not meet filter criteria for further processing or do not have sufficient information for a stage are diverted. Once an annotated source reaches the bottom of the \texttt{ANTARES} architecture it is stored in our locus-aggregated alert database, made publicly available, as well as broadcast to external alert-brokers. Additionally, we will filter out the most rare alerts, broadcasting them separately to facilitate coordinated followup studies across the electromagnetic spectrum.

In \S\ref{sec:overview}, we stated three questions to motivate the development of stages for the pipeline in this work: how effective is machine learning at a) early-time categorization of variables and transients, b) classification into broadly separable astrophysical classes without full phase curve information, and c) late-time retrospective classification aimed at producing a high-purity sample of objects. Each of these questions led to the construction, in \S\ref{sec:featureextraction}, of a stage to encode the information contained in the light curve as a low-dimensional feature vector. Having computed the matrix of feature vectors for the dataset, we then define filters to \textit{select} alerts and train machine learning algorithms to \textit{categorize} and \textit{classify} them -- examples of core machine-learning stages that will comprise the \texttt{ANTARES} pipeline. These stages parallel our motivating questions of machine-based selection, categorization and classification. 

We plan to implement iterated semi-supervised learning by using this pipeline to label new datasets, and then using those sets to retrain our classifiers. As we do so, the stages we use will undoubtedly evolve significantly. This iterative process of machine learning, classifier retraining, and stage modification will continue throughout the entire lifetime of the LSST project. Consequently, brokers will need to adopt version control, not only for their code base, but also to track the provenance of the library of datasets, the feature matrices, the different splits used for training and testing, and binary representations of the filtering and classification stages: the ``Touchstone'' illustrated in Fig~\ref{fig:architecture}. We structure the different stages hierarchically, to reflect how they are used to process sources with different amounts of phase coverage, hence with different amounts of information encoded in the feature vector. 

\subsection{Putting it All Together: Constructing an Alert-Broker Pipeline}

\begin{figure*}[htpb]
 \vspace{1em}
 \centering
 \includegraphics[width=\textwidth]{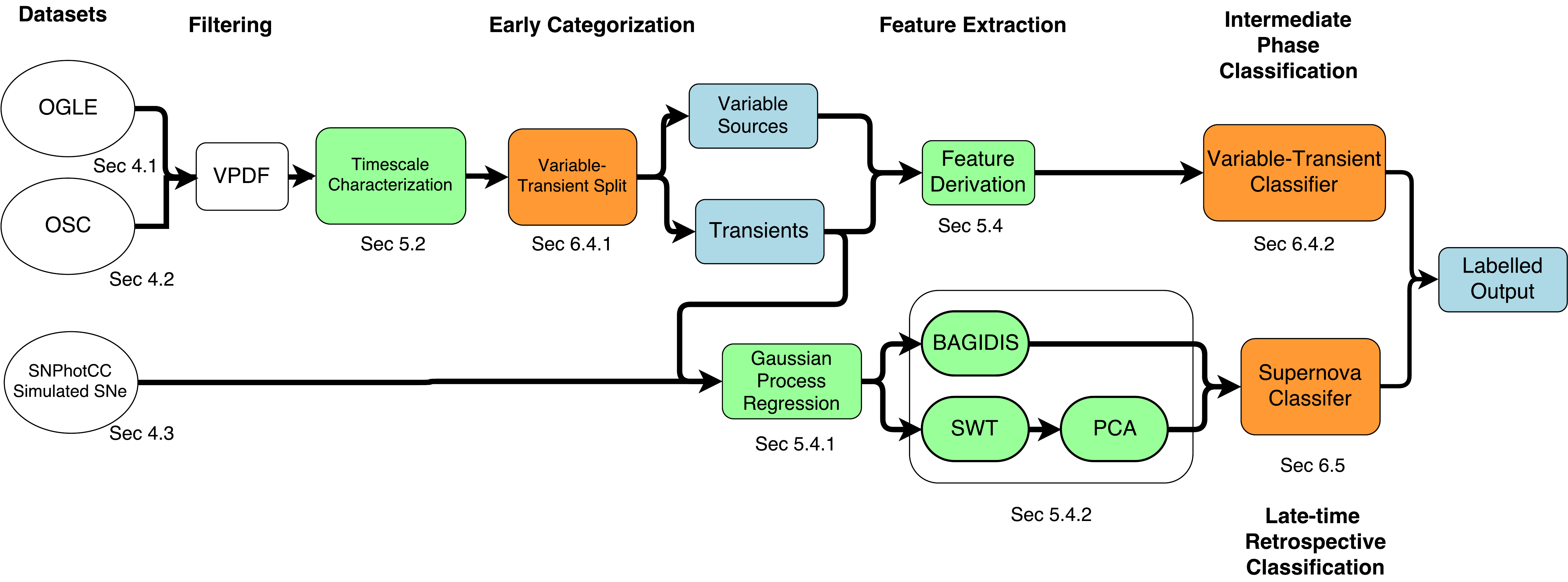}
 \caption{Schematic of the \texttt{ANTARES} machine learning pipeline, corresponding to the yellow bracketed region in Fig.~\ref{fig:architecture}. The pipeline stages are designed to examine three different use cases for an alert-broker, corresponding to the different amount of data in the alert packets for each source, as described in \S\ref{sec:overview}. The input datasets are described in \S\ref{sec:datasources}, the feature extraction in \S\ref{sec:featureextraction} and the various stages in \S\ref{sec:machinelearningpipe}. The specific subsections in this work corresponding to each element of the pipeline are indicated below the stage. Stages that extract feature information are indicated in green, categorization or classification stages are indicated in orange, and outputs are colored in blue.}
 \label{fig:pipeline}
\end{figure*}

When an alert is received by the pipeline developed for this work, the first processing stage consists of a filter using the most basic features computed in \S\ref{sec:vpdf}. This stage is filtering as opposed to classification, and is constructed from the best available unbiased catalog of stellar variability available -- at present, the \textit{Kepler} sample. Consequently, it involves no machine-learning, and we do not consider it further in this section.

As additional observations of the target are acquired, we can perform more advanced characterization of the time-series, as described in \S\ref{sec:timescalefeature} and \S\ref{sec:magcharacterization}, and attempt classification. We expect that stages that operate in this intermediate regime, where some observations have been acquired after the initial alert but the full phase curve of the event has not been probed, will serve the bulk of the astronomical community's needs. 

To begin with, we attempt to determine whether the object is a variable or transient. The only transient type considered here are supernovae, though LSST will produce alerts from other astrophysical classes of transients. This is not a significant limitation for distinguishing between variable and transient sources, and the choice to use only supernovae in this work reflects the labelled datasets that are readily available for training. As we use this work for semi-supervised learning with Pan-STARRS, we will be able to construct a much more homogeneous training set. We do not expect stages like this to become substantially more complex, as they are ultimately doing binary classification. 

The next stage after characterizing the alert and determining if the object is a recurring variable or transient is to attempt to determine its class. This stage is intrinsically more complex than variable--transient separation, as it involves multi-class labelling; a problem that is exacerbated by the extreme population imbalances in our combined OSC and OGLE samples. Even with a more homogeneous data sample drawn from a single survey, this imbalance will persist as different classes of objects have very different astrophysical rates, and therefore any classifier attempting to tackle this problem must adopt strategies to deal with unbalanced, and likely non-representative, training sets. 

Finally, once enough of the time-evolution of the source has been probed, we can perform complex feature derivation, such as the quantities extracted by the processes described in \S\ref{sec:advancedcharacterize}. While this processing is typically CPU intensive, it need not be repeated for transient objects, and is unlikely to need repeating for variable sources until a sufficiently long time baseline is covered by the observations, which would allow us to look for long-term variability. In this regime, we look at a specific use case of scientific interest to many groups -- the extraction of a high-purity photometric sample of SNIa for cosmological investigations.

The pipeline and stages developed for this work are represented graphically in Figure \ref{fig:pipeline}. Each of the stages of our machine-learning pipeline requires the training and validation of a classification algorithm. We use a random forest algorithm for all of the machine learning tasks in this work. While there are several alternative machine learning algorithms we could employ, random forests have several advantages,described in the following section, that make them particularly suitable for this study. 

\subsection{Decision Tree Learning \& Random Forests}\label{sec:randomforest}

Decision tree learning allows for a mapping from input features to output classes by means of a series of selection rules. An individual tree is trained by generating a selection rule based upon whichever feature and threshold gives the maximum information gain, where information gain can be determined by several different metrics. The two most common choices are the Gini-impurity:
\begin{equation}
\label{eq:gini_impurity}
   I_{G} (p) = \sum \limits_{i \ne k} p_i p_k 
\end{equation}
or the entropy: 
\begin{equation}
\label{eq:entropy}
   I_{E} (p) = \sum \limits_{i = 1}^{J} p_i \log_{2} p_i 
\end{equation}
where $p_i$ is the percentage of the $i$th class in the sample of $J$ classes present at the child node of each split. The percentages are normalized to sum to unity. The information gain is defined as the difference between the metric computed for the parent node the weighted mean of the metric for the child nodes (i.e. after the selection rule is imposed). The generation of selection rules proceeds recursively until objects in the training set in a given area of a tree are all of a single class, or until a given threshold of tree depth. 

An individual decision tree classifier that is grown ``deep'' is typically overfit, i.e. has a very high variance. The opposite is true for a decision tree classifier that is grown ``shallow.'' It will have low variance, but high bias, i.e. is underfit, and does not capture the relationships between the feature vectors and target outputs. Using a single decision tree for classification generally yields poor performance, so decision trees are rarely used on their own for machine learning \citep{james2013}. Instead, powerful ensemble methods have been developed that rely upon the averaging of errors between ensembles of trees. A popularly-used method is random forests \citep{breiman1999}.

A random forest classifier is a learning method that trains a ensemble of decision tree classifiers and takes the mode of the classification results as the output. Random forests have been used to great effect in a number of fields, and in particular for photometric supernova classification \citep{Lochner2016}, but also in other areas of astronomy \citep{2015Carrasco, 2011Dubath}. L16 and others have demonstrated that random forests show similar performance to other classification algorithms in the context of astronomical time-series datasets. Therefore we employ random forests for all three learning tasks -- early variable/transient categorization, intermediate variable and transient classification, and late-time SNIa/non-Ia separation -- that we consider in this work. 

The use of an ensemble of decision trees in the random forest has the effect of greatly decreasing the variance of the classifier, but increases the bias. The algorithm samples the instances in the training set repeatedly with replacement (randomly in the case of a random forest). If there are $N$ features for each instance, a threshold can be specified of $n$ such that $n\ll N$. Only $n$ variables randomly selected from the $N$ features are then used when growing the individual decision tree. Each tree is grown as far as possible until all objects in the end branches of a tree are of a single class. When seeking to classify a new object, the input features are processed by all the trees in the forest. Each tree outputs a classification based upon its selection rules, and ``votes'' for the class it determined. The mode of the class selections determines the output classification of the forest. By aggregating ensembles of decision trees, random forests avoid the bias-variance trade off that is inevitable with a single decision tree. 

Random forests also provide several useful metrics ``for free'': the out-of-bag (OOB) decision function, and relative feature importance. For each random subset of the data selected in the first step of training, there is a portion of the data that was not used in the building of the decision tree (the out-of-bag data). For any given object, there will be approximately one-third of the total number of trees that never used the object in the tree generation \citep{breiman1999}. The decision function is then the classification result from running the each of the OOB objects through the two-thirds of the trees that did not use them for training. An error estimate is generated by comparing the decision function to the labelled classifications. Random forests also provide a natural robust estimate of the feature importance. At each successive split made on a given feature, $m$, when training a single decision tree, the algorithm computes the decrease in the weighted impurity. For a forest of decision trees, the weighted impurity decrease for each feature can be averaged across all trees, and the features ranked by this average. 

Random forests effectively memorize the data used for training. For a thorough analysis of generalization error, it is essential that either independent data be used for testing, or the OOB error estimate can be taken. The OOB error estimate has been shown to be a biased estimate of the generalization error and an external testing set is still important to validate results \cite{bylander_estimating_2002}. We use the \texttt{RandomForestClassifier} implemented by the \texttt{scikit-learn} python package\footnote{\url{http://scikit-learn.org/stable/}} \citep[][]{scikit-learn} throughout our pipeline. We describe the development and training of each of the stages of our machine learning pipeline in the following sections.

\subsection{Classification of Variables \& Transients}\label{sec:vartrans}

After LSST issues an alert, it will continue to monitor the location of the source on all subsequent visits. While we initially use contextual information about the source as well as filtering stages such as the VPDF (\S\ref{sec:vpdf}), as more of the time-evolution of the source is observed, we can perform increasingly complex characterization of the physical timescales and flux changes of the event. With this additional information on each source, we can use machine learning techniques for better categorization and to attempt classification with finer sub typing. The most basic binary decision that can be made is if a source is a variable or a transient, whereas the VPDF only determines if the variability is significant with respect to the local stellar background.

\subsubsection{Variable--Transient Separation}

For variable vs transient separation, we assemble a vector of the features computed in \S\ref{sec:timescalefeature} for each of the two passbands ($g$ and $i$) for each object in the combined OSC and OGLE sample. The advantage of constructing the vector from only timescale features is that such information is frequently reported in external catalogs of variable sources such as GAIA and Pan-STARRS. This will allow us to populate the feature vector, even in the absence of many LSST observations, using surveys which observe in a different set of band passes.

All objects in the OSC sample were considered transient, irrespective of sub-type, whereas all objects in the OGLE sample were considered to be recurring variables. For both the OSC and the OGLE data, only the $g$ and $i$ passbands were taken after passband mapping. If they were not present the light curve was discarded. As \citet{Kim2015Upsilon} note, several period-related features such as period $S/N$ ratio, the period itself and associated uncertainty allow for distinguishing between periodic and non-periodic objects. Requiring timescale information from two bands increases our sensitivity to this split, as periodic variables often exhibit similar behavior in different passbands, whereas transient sources may not. The OGLE objects generally have an observation period that extends far longer than that of the OSC SNe, so we avoid using features that operate as a proxy for the duration of observation. This is a limitation that arises from the heterogeneity of our sample. We expect to be able to derive more complex features for this stage when operating on homogeneous datasets.

To classify the OGLE and OSC light curves as recurring variables or transients, we used a random forest classifier with balanced class weights. This re-balancing helps to account for an $\sim$8 to 1 ratio of variable to transient objects in the sample. The re-balancing adopted here is simple: the random forest applies a weight to the minority class in inverse proportion to its percentage of the training data. These weights are used in the development of the decision tree in two places -- in the weighting of the Gini impurity coefficient, and in the final vote tallying for an object; the decision becomes a weighted majority vote in accordance with the balanced class weights as opposed to a simple majority. There are many alternative approaches to rebalancing the training sample, and we will examine some in other stages of our pipeline. However, for this learning task, we can use cross-validation to show that the simple weighting approach suffices. 

To evaluate the consistency of the classifier, we calculate the evaluation scores taking the average over a 5-fold cross-validation. The data was split up into five different training and testing sets, or ``folds''. Each fold's training set was used exactly once, with the remaining data used for testing in each iteration. The classifier was run using 200 decision trees, though the classification performance was not highly dependent on the number of trees chosen, as we expect for a simple binary decision task. For the subsequent stages in the pipeline where we consider increasingly complex problems, we adopt increasingly sophisticated approaches to tuning the classifier. We report the results of this classification stage in \S\ref{sec:classifierperformance}

\subsubsection{Variable \& Transient Classification}

While variable--transient separation is of utility for many follow up studies, a labelled feed of alerts with a high confidence of belonging to a particular astrophysical class is very desirable and serves a large range of scientific interests. As the full time-evolution of the sources has not been probed, the input features must be chosen carefully to be robust to outlying photometry, and stabilize as more data is added. If this condition is not satisfied, it is likely that the predicted classification will not be stable; i.e. it will change frequently as more observations are added. We construct the feature vector for each source following \S\ref{sec:magcharacterization}.

For the binary classification problem of variable-transient separation, class labels are aggregated into either variable or transient for the training, and consequently the class imbalance of the combined OGLE and OSC dataset is much reduced. However, as we are now attempting to classify variables and transients into their sub types, we cannot aggregate labels. This fundamentally multi-class problem requires that we contend with the extreme class imbalance in our data sample. We account for the class imbalance using a combination of techniques; under-sampling the majority class, aggregating minority classes that are very similar into super classes, and synthetic minority oversampling. 

\paragraph{Under sampling and Aggregation to Reduce Class Imbalance}\label{sec:resample}
With $\mathcal{O}(10^6)$~members, the long period variables in the OGLE dataset outnumber every other class in the combined sample of OGLE and OSC objects. Therefore, we under-sample this class by using 10\% of the total samples for training and testing, reserving the rest for validation. This makes the number of available long period variables comparable to the next biggest classes in the OGLE sample -- RR Lyrae and eclipsing binaries. 

Additionally, we can aggregate some classes that share many similarities together in the training sample. For this work, we elected to combine eclipsing binaries and ellipsoidal/contact binaries together under the Algols label, and we combined classical Cepheids, double-mode Cepheids and type II Cepheids together under the ``Cepheid Variables'' label. 

This aggregation is reasonable given the design goals of this stage, which we expect will process $\sim 10$ observations from LSST for each source. This is unlikely to be sufficient to distinguish between the aggregated sub types; high-confidence classification into such sub types will require a long baseline of observations to constrain variability on extended timescales. 

While these two methods help address some of the class imbalance, they are insufficient by themselves, as many of the classes of variables still outnumber the transients by two orders of magnitude. The class distribution of the full dataset after undersampling and aggregation is listed in Table~\ref{tab:var_trans_RF_sample_size}. 

\begin{table}[ht]
 \centering
 \begin{tabular}{p{1.5in}|c}
 \toprule
 Class      & Number of Objects \\ \tableline
 Algols      & 39,452 \\
 RR Lyrae  & 38,243 \\
 Long Period Variables & 28,904 \\
 Cepheid Variables   & 8,672 \\
 $\delta$ Scuti & 2,844       \\
 Supernovae & 1,048   \\ 
 Double Period Variables & 136 \\ \tableline
 \end{tabular}
 \caption{Sample sizes of the classes in the combined OSC and OGLE dataset after feature extraction from light curves with a sufficient number of observations, and after undersampling of long period variables and aggregation. Despite these techniques, the sample remains extremely imbalanced with classes of interest, such as supernovae, being underrepresented relative to variables by two orders of magnitude. We use SVM-SMOTE (\S\ref{sec:smote}) to generate synthetic samples from minority classes, and improve the classifier's sensitivity to class boundaries.}\label{tab:var_trans_RF_sample_size}
\end{table}

\paragraph{Dealing with Extreme Class Imbalance: Synthetic Minority Oversampling}\label{sec:smote}

Random forests include the ability to account for class imbalance using re-weighting or random sampling with replacement. Randomized oversampling duplicates members of the minority class in the training set, thereby preventing classifiers from being dominated by the majority class. However, this technique is naive when the class imbalance is drastic, as in the case of the combined OGLE and OSC dataset, with some classes having 2--3 orders of magnitude more members than others. Each member of the minority class in the training set is duplicated several times, and we found unsurprisingly that this leads machine learning classifiers to overfit the duplicated samples at the expense of precision and recall when validated with test samples. Therefore we employ the Synthetic Minority Oversampling Technique \citep[SMOTE,][]{SMOTE} to \emph{generate} new samples of the minority class, rather than simply duplicate existing members. 

SMOTE is fundamentally a multi-dimensional interpolation scheme. The standard implementation of the algorithm first constructs a k-D tree of the training set, and then determines which classes are in the minority and require over-sampling. When resampling to create a new synthetic data point, the algorithm selects one of the k-nearest neighbors in the feature space of a selected minority class member, and constructs the vector between the selected point and its neighbor. The algorithm draws a random number between 0 and 1, and multiplies the vector by this number to generate a new point. 

The basic implementation of SMOTE can be affected by outliers in the minority class, causing the algorithm to extrapolate new points in regions of the feature space where the minority class is not otherwise represented. We therefore employ the more sophisticated SVM-SMOTE \citep{SVMSMOTE} variant of the algorithm. This variant constructs a Support Vector Machine (SVM) to determine boundaries between classes, and interpolates new members near the border lines, to increase the machine learning classifier's sensitivity to class separation. We utilize the SVM-SMOTE implementation from the \texttt{imbalanced-learn}\footnote{\url{http://contrib.scikit-learn.org/imbalanced-learn/stable/index.html}} \citep[][]{imbalancedlearn} package in our pipeline. 

We emphasize that it is critical that SMOTE and its variants be applied \emph{only} to the training set, and not the full dataset prior to the train/test split; the latter would only result in overfitting of the interpolated samples rather than producing any improvement by accounting for the imbalance between the classes. There are even more sophisticated techniques that combine oversampling methods such as SMOTE with undersampling techniques like Edited Nearest Neighbors (ENN), to ensure the resampled training sets are free of noise. \citet{resampling} compares several methods for rebalancing, and we expect many of these techniques to become more prevalent in the domain of astrophysics as groups are forced to contend with the imbalanced samples that will be produced by wide-field synoptic surveys.

However, even minority oversampling techniques such as SVM-SMOTE may be insufficient to address classes that are completely underrepresented, and that have a high dispersion in feature space. We have deliberately included the double period variables with only 136 members in the sample to illustrate this. The class consists of objects that show more than one fundamental period, but do not appear to belong to other astrophysical classes. The class itself then is ill-defined, with members sharing properties with other classes, and consequently having a large dispersion in feature space. This makes construction of class boundaries using SVM sensitive to the specific objects included in each of the k-fold training samples. The naive use of resampling techniques in this scenario may lead to biased classifiers that capture spurious relationships generated by interpolation and the target outputs. Astronomers employing such imbalanced learning techniques must make an informed choice as to \emph{how} to rebalance the training sample, rather than employing these sophisticated algorithms as black boxes.

\paragraph{Training \& Cross Validation}

\begin{figure}
  \centering
  \hspace{-20pt}
  \includegraphics[width=0.5\textwidth]{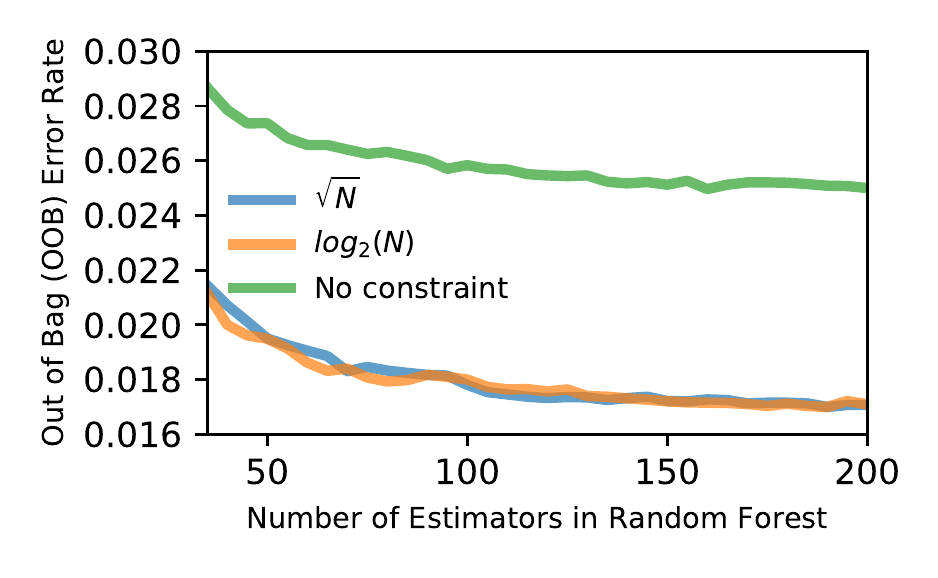}
  \caption{Tuning a random forest by examining Out of Bag (OOB) error rate vs number of decision trees (estimators). The curves correspond to different constraints on the maximum number of features that each tree can consider from our $N=15$ dimensional feature vector.} 
  \label{fig:RFTuning}
\end{figure}

We use 9-fold cross-validation, splitting the dataset of $N$=15 dimensional feature vectors (see \S\ref{sec:magcharacterization}) into nine different training and testing samples. We reserve 60\% of the data for testing, and use the complement for training. We apply SVM-SMOTE to each of the 9-fold samples to generate a new balanced training set for machine learning with the random forest algorithm. We tuned the algorithm by examining the OOB error rate for random forests with a different numbers of trees (see Fig.~\ref{fig:RFTuning}), and selected an ensemble size of a 100 decision trees; increasing the number further only results in a marginal decrease in the OOB error and increase in the accuracy. 

We also examined the OOB error rate vs the maximum number of allowed features for classification, testing the behavior if we imposed no constraint, or limited the maximum number of features to at most $log_2(N)$ or $\sqrt N$. We found little difference between the two methods of restricting the number of features. However, both of these methods consistently outperformed no constraint whatsoever, at all ensemble sizes. Closer examination of the outputs from the classifier suggests that a lack of constraint allows the decision trees in the ensemble to adopt selection rules based on features that produce very little change in the entropy, whereas restricting the maximum number constrains the trees to select features that maximize the Kullback-Leibler divergence, i.e. the information gain. This behavior is exactly analogous to the optimization step of the t-SNE visualization (\S\ref{sec:class_imbalance}). We report on the results of the classifier in \S\ref{sec:classifierperformance} using a forest with 100 trees, trained to optimize the information gain, and with the maximum number of features limited to $log_2(N)$.

\subsection{Supernova Classification}\label{sec:sne_classif}

Acquiring enough observations to cover the full time-evolution of the sources necessarily means that classification for transient objects is retrospective. However, this regime is still of interest for several groups engaged in population studies. One of the biggest challenges of these studies will be discriminating between the class of interest and ``impostors'' -- objects that have very similar light curves, but different underlying astrophysics. This is the key distinction to the stages described in \S\ref{sec:vartrans} that are designed to separate sources into broadly distinct astrophysical classes using features derived from only a section of the light curve. Brokers like \texttt{ANTARES} must extract the maximum amount of information available in the light curve and build complex classifiers that are capable of discriminating between the class of interest to meet the requirement for a high purity photometrically selected sample.

The specific use case we consider in this section is the need for a stage capable of producing a photometric sample of SNIa for a cosmological study to constrain any evolution of the equation of state of the dark energy. Several groups have considered this problem in detail since SNPhotCC became available, offering a benchmark dataset and a rich literature against which we can compare results. We adopt a variant of the wavelet-based classification methods employed in L16 to extract information from the SNe light curves. 

Non-parametric representations of the time-evolution have a distinct advantage over template-based methods for alert-brokers: the same features can be effectively extracted from most of the light curves, whereas template fitting or parametric representations are typically tuned to at most a few classes of astrophysical objects. We construct the feature vector by modeling each light curve with a Gaussian process (\S\ref{sec:gpmodeling}), and applying wavelet transformations (\S\ref{sec:wavelet_transforms}) to the interpolated output. While this processing is considerably more computationally intensive than deriving the feature vector for variable--transient classification (\S\ref{sec:vartrans}), because this stage is run on the output of the previous stage, it only has to consider the fraction of the alerts that have a high confidence of not being a recurring variable. This choice reflects a simple design philosophy that we have adopted for \texttt{ANTARES}: ``do the least with the most, and the most with the least" -- i.e. we apply expensive analyses only when the percentage of relevant alerts is low.

As we are largely comparing objects with multiband photometry from the OSC and SNPhotCC against each other, rather than against OGLE, we use all of the available observer frame bands that can be mapped to the $griz$ as described in \S\ref{sec:bandmapping}. Portions of the data were removed from the classification process due to selection cuts. For the SNPhotCC, if the objects did not have coverage in all four of the \textit{griz} filters, the light curve was discarded. Out of the 17133 light curves in the dataset, 279 were eliminated by this selection cut. For the OSC, the criterion was relaxed, and only objects with observations in $g$, $r$, and $i$ filters were used for classification. Of the 3259 light curves, 2035 were eliminated by this selection cut. Using the same selection requirements as the SNPhotCC for the OSC would have resulted in a wholly unacceptable loss of 3,118 light curves, leaving only 141 light curves for analysis. While we could relax this further to only require a single color, and thereby include more objects, this is not reflective of the data that LSST will produce, or the existing Pan-STARRS and the upcoming simulated PLAsTiCC datasets to which we wish to adapt this pipeline, defeating the goals of this work. 

\begin{figure}
  \centering
  \includegraphics[width=0.4\textwidth]{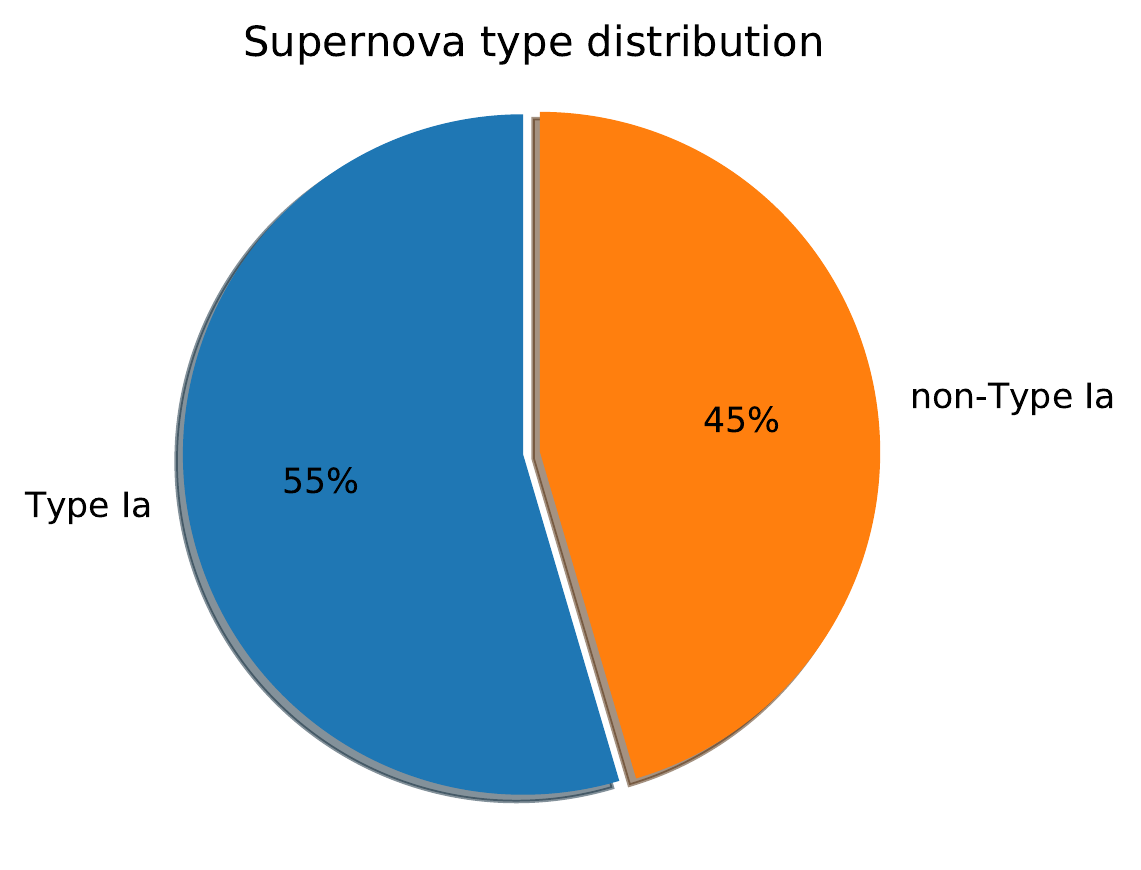}
  \caption{Distribution of SNe in the OSC after aggregation into binary Ia/non-Ia labels. This aggregation is necessary as many SNe subtypes are not distinct photometrically. The resulting dataset is not strongly imbalanced.}
  \label{fig:snbin}
\end{figure}

We reduced this stage to a binary classification system, using type Ia supernovae (including subtypes) as the positive class and non-Ia as the negative class. The non-Ia supernovae included both Ib/c and II when classifying both the SNPhotCC and the OSC data. This aggregation into binary labels is unfortunate but necessary given that the labels in the OSC dataset (see Fig.~\ref{fig:classdistrib}) are determined by spectroscopic indicators, and many of the subtypes are not substantially distinct photometrically (e.g. Ia-91T, Ia, Ia-91bg, Ia-02cx). Taming the supernova zoo with aggregation has the bonus of producing a relatively balanced dataset (Fig.~\ref{fig:snbin}), allowing us to employ random sampling with replacement, rather than the more complex SMOTE approach.

\subsubsection{Training \& Cross-validation with Hyperparameter Optimization}\label{sec:hyperopt}

All machine learning algorithms have several hyperparameters that require optimization. For this work, many of the defaults for the random forest were considered to be appropriate for initial examination, with only the number of estimators taken as a potential variable. This is suitable for most of the classifiers, which only utilize a small number of features. However, our wavelet feature extraction stages produce a very large number of feature coefficients ($\mathcal{O}(10^{2})$ for the SWT). 

While we employ dimensionality reduction, either in the form of principal component analysis or the BAGIDIS decomposition to reduce the number of coefficients, the wavelet transformation still results in an extremely high dimensional feature space. Optimizing classifier performance (and therefore the purity of the extracted SNIa sample) requires that we investigate how many, and what features from this extremely high dimensional space are useful. Therefore, we optimize the number of wavelet coefficients $k$ and the maximum number of decision trees allowed for classification by the random forest, $N$, for this stage of the pipeline.

\begin{table}[htpb]
  \subfloat[SNPhotCC]{
  \centering
  \begin{tabular}{p{1.3in}|c|c}
    \toprule
    Wavelet Scheme  & $k$ & $\sigma$ \\ \tableline
    Daubechies & 112      & 9.6      \\
    Symlets  & 109      & 9.7      \\
    BAGIDIS  & 8      & 0.48      \\ \tableline
  \end{tabular}
  \hfill
  }
  \hspace{0.2cm}
  \subfloat[Open Supernova Catalog]{
  \centering
  \begin{tabular}{p{1.3in}|c|c}
    \toprule
    Wavelet Scheme & $k$ & $\sigma$ \\ \tableline
    Daubechies & 66      & 16      \\
    Symlets  & 66      & 13.6      \\
    BAGIDIS  & 9      & 0.6      \\ \tableline
  \end{tabular}
    \hfill
  }
  \caption{The optimized hyperparameter of the number of wavelet coefficients, $k$, used in classification of the SNPhotCC (top) and OSC (bottom) datasets. These numbers are averages over five iterations to assess model stability and the standard deviations of each value are on the right.}
  \label{tab:hyperparams}
\end{table}

To assess model stability for the SNe classification, we used 5-fold \emph{nested} cross validation. Nested cross validation allows for the optimization of hyperparameters combined with evaluation on a separate hold-out set. This ensures that the optimization step does not bias the classifier and allows for the entirety of the data to be used for training and testing once all iterations are completed. 

The classifier optimization step is nested within the hyperparameter optimization. We ran a hyperparameter search for each cross-validation fold of the SNe classifier. This optimization routine maximized the discrimination between the correctly an incorrectly predicted inputs from the classifier by varying the number of input components returned by the dimensionality reduction methods, $k$, listed above. The results of this randomized search are listed in Table \ref{tab:hyperparams}. 

The hyperparameter for the number of wavelet coefficients used in classification, $k$, was stable over the five iterations for all wavelet types. As expected by their design, the BAGIDIS decomposition coefficients required far fewer components for peak classification performance than the Daubechies or symlets coefficients. The SNPhotCC in general required a larger number of principal components when maximizing performance.

Neither the OSC or the SNPhotCC depended strongly on the number of decision trees used in the forest, $N$, as long as the number was reasonably high ($N \ge 300$). This is in keeping with other work and our finding in Fig.~\ref{fig:RFTuning} that suggest that as long as the number of trees in a given random forest is above a certain threshold, the performance increases only slightly with more trees \citep{breiman1999}. We utilized a forest with 600 trees for analysis of all the different wavelet methods, for both the OSC and SNPhotCC data. 

\section{Machine Learning Classifier Performance}\label{sec:classifierperformance}

After training each stage of our pipeline, we evaluated its performance with cross-validation on test sets. We use several standard statistical quantities to assess the stages, and we briefly describe these evaluation metrics below. \\ \\

\subsection{Evaluation Metrics}

We use three metrics for evaluation of the classifier performance: 1) accuracy, 2) the receiver operating characteristic curve and 3) the normalized confusion matrix. All three metrics have been used previously in the astronomical literature on machine-learning (for e.g. R11 and L16). We briefly describe the properties of these metric below.

Most evaluation metrics are defined in terms of binary classification, with one class being considered a ``positive'', and the other being ``negative''. The basic quantities used to describe the input sample are the number of positive and negative cases in the sample, $P$ and $N$. The quantities used to describe the classified sample are the number of correctly classified ($T$) objects -- the ``True Positives'', $TP$, and ``True Negatives'', $TN$. Their complement is the number of incorrectly classified ($F$) objects -- the ``False positives'', $FP$ and ``False negatives'', $FN$. These quantities can also be used in a multi-class scenario by taking a ``One-vs-rest'' approach, where one class of the sample with $J$ classes is treated as positive, and all the others are treated as negatives. 

\subsubsection{Accuracy}
The simplest metric that is used is the accuracy with which a trained classifier predicts the labels of a test set -- the fraction of correctly predicted labels:
\begin{equation}
 \text{Accuracy} = \frac{TP + TN}{P + N}
\end{equation}

Accuracy is a poor metric of evaluation if the classes are not evenly distributed, as is the case for the SNPhotCC dataset (30\% Type Ia). In such cases, several other metrics can be used to evaluate classifier performance while accounting for class imbalance, and these provide more useful measures of classifier performance in astrophysical contexts. We evaluate the receiver operating characteristic (ROC) curve and the area under the curve (AUC), in this work. 

\subsubsection{Receiver Operating Characteristic Curves}

Receiver Operating Characteristic (ROC) curves visualize the trade-off between sample purity and sample completeness -- i.e. the true positive rate (TPR), and the false positive rate (FPR) as a function of the threshold used for classification. The true positive rate, $TPR$, is the ratio of correctly classified positives to the total number of positives in the dataset:
\begin{equation}
 TPR = \frac{TP}{TP + FN} = \frac{TP}{P}
\end{equation}
and similarly, the false positive rate, $FPR$, is defined as:
\begin{equation}
 FPR = \frac{FP}{FP + TN} = \frac{FP}{N}
\end{equation}

The output classification probabilities from the classifier are on the unit interval. If the threshold for a positive classification is set to 0.8, the classifier will not classify the object as a positive until the probability exceeds 80\%. By varying this threshold parameter, we can better assess the performance of the machine learning algorithm. The ROC curve varies the threshold continuously, evaluating the TPR and FPR for each value of the threshold. This additionally provides a metric to determine what threshold is appropriate for a study. Studies that understand the impact of false positives, such as cosmological studies using photometric samples of SNIa, can determine a threshold that provides the desired sample purity. We use the ROC curve to evaluate the performance of our wavelet-based SNIa classifier for the OSC and SNPhotCC datasets in \S\ref{sec:waveletresults}.

The goal for classification is to maximize the TPR, while simultaneously minimizing the FPR. A useful single-number metric that can be extracted from an ROC curve is the fractional area under the ROC curve (AUC). If the AUC is equal to 0.5, then the TPR equals the FPR for all thresholds, indicating random classification, i.e. an uninformative ``guess''. An AUC of 1 indicates that the TPR is maximized at all thresholds and the FPR is minimized, representing perfect classification. 

\subsubsection{The Confusion Matrix}

While the ROC curve is a useful metric to study classifier performance in a ``One-vs-rest'' context, it is often useful to know what the false positives are - i.e. which astrophysical sources are the source of contamination for the class of interest. The normalized confusion matrix evaluates the fraction of each input class as a function of each output class:
\begin{equation}
 c_{ij} = \frac{N_{ij}}{P_i} \\
\end{equation}
where $N_{ij}$ is the number of elements of class $i$ labelled as class $j$. Where a visualization such as the t-SNE (\S\ref{sec:class_imbalance}) can provide a \emph{qualitative} assessment of what the sources of contamination will be prior to machine learning, the confusion matrix provides the quantitative assessment after the classifier has been developed. Both approaches have merit: the first aids in the design of the feature encoding and the classifier, while the second provides a metric that can be used to compare different classifiers against each other. We use the confusion matrix to evaluate the performance of our variable and transient classification for the combined OGLE and OSC dataset in \S\ref{sec:vartransresults}.

\subsection{Quantified Classifier Performance}

In the following sections, we measure the various evaluation metrics for each stage of our pipeline, and compare to the literature wherever possible. While we discuss the performance of each stage separately, brokers will structure several machine learning algorithms into a pipeline, and the stages cannot be considered independently of each other. Simulated datasets such as the upcoming PLAsTiCC will be essential to evaluate broker performance.

\subsubsection{The Variable--Transient Split Using Timescale Characterization}

As discussed in \S\ref{sec:resample} and \S\ref{sec:smote}, real astrophysical datasets have complex selection effects that result in imbalanced class representation. Even in volume limited surveys, the fundamental differences in event rates can create an almost an order of magnitude imbalance, e.g. between recurring variables and transients. The ubiquitous class imbalance makes the simple accuracy score a poor metric for evaluation. We therefore construct the ROC curve for this stage, and evaluate the AUC. The periodic vs non-periodic classification performed well, with a consistent AUC of 0.99 when run over 5-fold cross validation. Only 8 per 40,000 objects are misclassified. 

LSST can expect this level of performance as early as a few months after operation. Most variables will be detectable in multiple passbands, and the survey will rapidly establish a baseline of variability for all sources, whereas transients will typically only have non-detections before explosion. Effectively then, this stage is distinguishing between a multiband Fourier transform of a continuous signal, and a multiband Fourier transform of a wave packet. In addition, we can expect external catalog information from surveys such as GAIA, Pan-STARRS and SkyMapper to all help distinguish between variable and transient sources.

The binary output assigned by this stage does not give us an indication of which classes of astrophysical variables are being erroneously labelled as transients. However we can get a sense for this from the next stage in the pipeline, discussed below, which attempts a broad classification of variable and transients using subclasses of major groups of variables.

\subsubsection{The Classification of the OGLE \& OSC Sample Using Timescale and Magnitude Features}\label{sec:vartransresults}

\begin{figure}[htpb]
    \centering
    \includegraphics[width=0.5\textwidth]{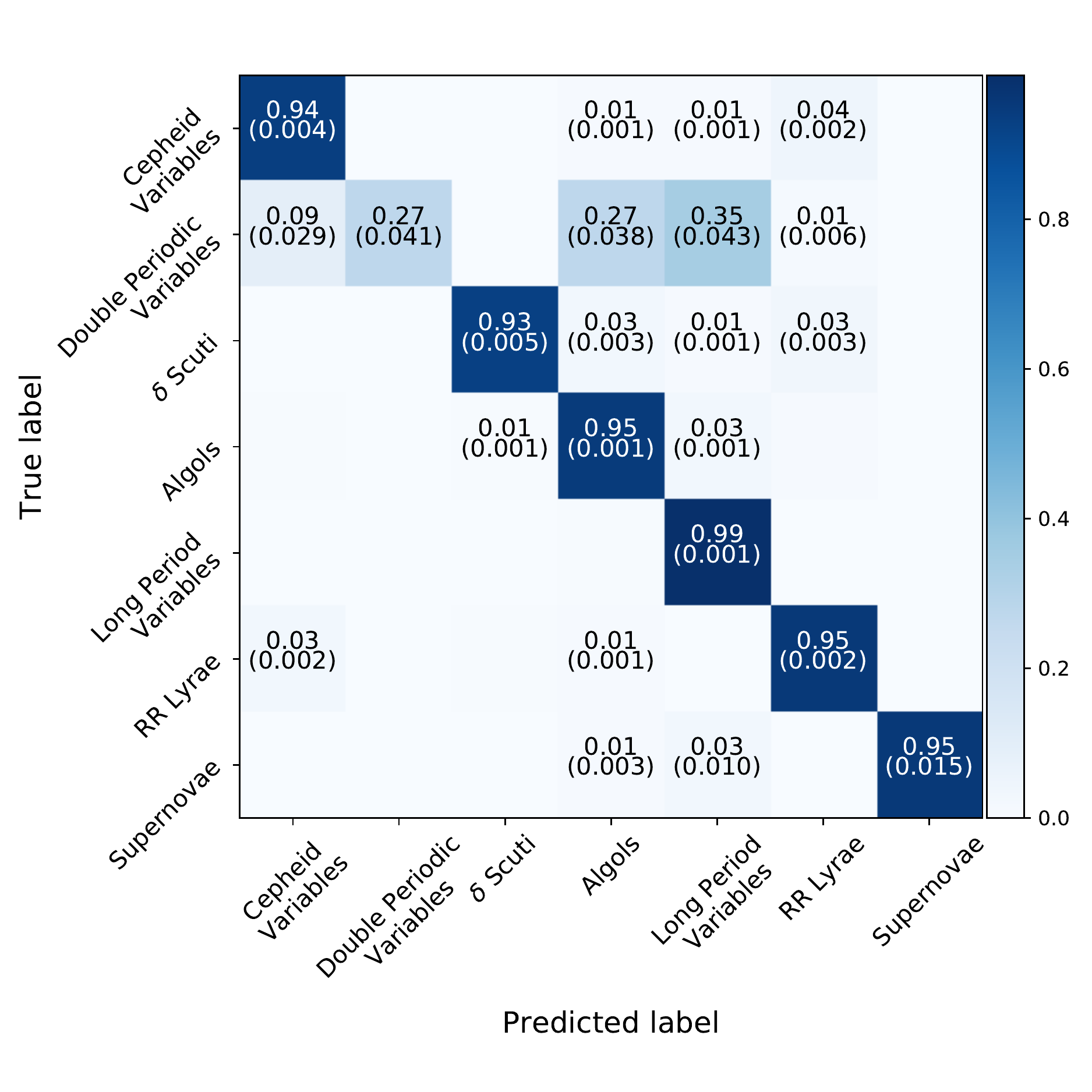}
    \caption{The normalized confusion matrix ($c_{ij}$) of the variable \& transient classifier applied to the 9-fold cross-validated test sets. We report the mean fraction of each of the true classes $i$ classified as a member of class $j$. The standard deviation of the confusion matrix elements is reported in parentheses. Elements without a numerical score have $< 1$~\% of the total class population. The only label on which the classifier performs poorly is the ill-defined, underrepresented class of double periodic variables.}
    \label{fig:OSC_OGLE_confusion_matrix}
\end{figure}

For non-binary classification problems, the receiver operator characteristic is not a curve, but a hyper-surface of every class against every other class, and is not easily interpretable. ROC curves can still be constructed by adopting a ``One-vs-Rest'' scheme, where one classifier is trained per class in the sample. However, this approach addresses a different question than originally posed, providing not multi-class classification, but a vector of non-comparable scores assessing the likelihood of a source being a member of each class, i.e. multi-label categorization. We therefore report the full normalized confusion matrix for the variable--transient classification stage, shown in Fig.~\ref{fig:OSC_OGLE_confusion_matrix}.

The overall accuracy of the classifier is 0.96. Despite the imbalanced dataset, all but one class in the set has an accuracy above 0.90, with most above 0.95. This indicates that the classifier is not overfitting the majority class, and the resampling approach we adopted in training (see \S\ref{sec:smote}) has been effective. The notable exception is the class of double periodic variables. This category is ill-defined and contains several different astrophysical classes, all of which exhibit at least two strong periods. Additionally, there are only 136 total members of this class, compared to $\mathcal{O}(10^{3}-10^{6})$ members in the other classes. The heterogeneity of the instances of this class, and their drastic underrepresentation, conspire to make any resampling scheme ineffective.

Nevertheless, such drastically underrepresented classes can be very interesting scientifically, e.g. the electromagnetic counterparts of gravitational wave sources such as GW170817 \citep{GWEMFollowup}. Alert-brokers will need to develop different strategies of identifying these extremely rare events in the data. It may be possible to use simulations to populate classes that are very rare, use techniques such as isolation forests to identify large outliers in the feature space, or develop bespoke filtering stages that include contextual information for these rare sources. But the overall performance of this stage augurs well for LSST, which will have more than the two passbands used for this study, and will be much more homogeneous and better calibrated. Multiband time-evolution with a cadence comparable to the characteristic timescales for many different kinds of sources will allow us to employ features that are better at discriminating between the different astrophysical classes and their subtypes.

\subsubsection{The Classification of Supernovae Using Wavelet Transformations of Light Curves}\label{sec:waveletresults}

\begin{figure*}[htpb]
  \centering
  \includegraphics[width=0.85\textwidth]{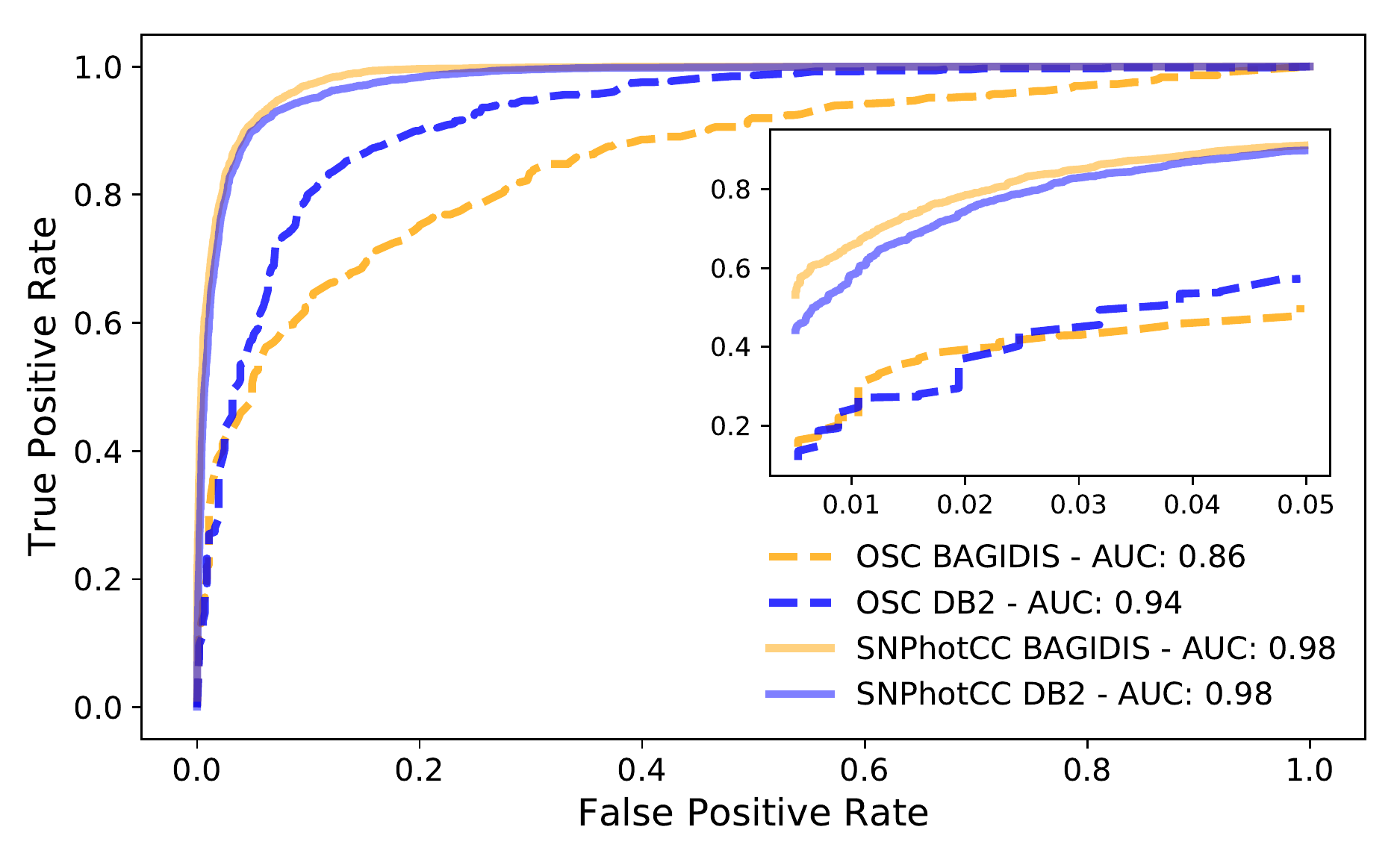}
  \caption{ROC curves for the SNPhotCC \& OSC datasets. The classifier performance drops on the OSC, relative to the SNPhotCC as indicated by the AUC score in the legend. This is illustrated in the inset, where the ROC curves are shown over a very narrow range in the FPR, necessary for a high purity sample of SNIa.}
  \label{fig:ROC_curves}
\end{figure*}

Supernova classification is the most complex stage of the pipeline. We are attempting to discriminate between the most-similar light curves of the dataset, to provide a high-purity feed to a cosmological experiment that is susceptible to bias arising from sample contamination. To accomplish this goal, we are using a large and difficult to compute feature vector together with the most carefully optimized machine-learning stage of our pipeline. Furthermore, as this stage is applied retrospectively, no additional observations can be obtained to improve the accuracy of the classifier. Figure \ref{fig:ROC_curves} shows ROC curves for the OSC and SNPhotCC datasets, for all wavelet classes and levels tested. 

\paragraph{Results for the SNPhotCC}
The AUC scores from the SNPhotCC show promising results, with values of 0.97--0.98 consistently for both the Daubechies and symlets features, and 0.98 for the BAGIDIS features. The SNPhotCC dataset has been used for testing classification schemes by many teams over the past seven years, and as a result there are many sources of comparison. The most directly comparable are those from \citet{Lochner2016}, as they used the AUC as their performance metric, as well as a similar classification scheme. Using Boosted Decision Trees, they achieved an AUC of 0.98 with symlets. An AUC of 0.98 reflects excellent classification performance on the part of our broker.

\paragraph{Results for the OSC}
This is the only work to date that has attempted to classify the heterogeneous and diverse set of all SNe publicly available through the OSC. There are no other figures of merit from other works against which to compare these results. Consequently, we compare the AUCs for the different wavelet decomposition methods against each other.

The BAGIDIS features perform significantly worse than both the Daubechies and the symlets, with an average AUC of 0.85, and particularly poorly compared to the SNPhotCC result. This is likely because the simulated SNPhotCC data are all generated from the same set of underlying models with the same survey characteristics and are much more homogeneous than the OSC data. The heterogeneity of the OSC data introduces noise into the BAGIDIS features, consequently decreasing classifier performance. For the OSC, there is no significant difference between the symlets and Daubechies wavelet results, both reporting high AUCs of 0.94. \\ 

\section{Discussion and Concluding Remarks}\label{sec:conclusions}

The \texttt{ANTARES} project described in this paper is a fully automated implementation of a machine-learning pipeline, forming the core of an alert-broker. 

This work serves two purposes. The first is to provide illustrative examples of the development and functioning of an alert-broker: what issues must be considered, what challenges arise, what techniques exist to address to mitigate them, what open-sources packages exist, and what level of performance is possible at present. 

The second purpose of this work is to develop a pipeline to process existing and upcoming datasets (such as Pan-STARRS and ZTF) through the \texttt{ANTARES} pipeline, in order to implement iterated semi-supervised learning. In an ideal world, we would already have a readily available representative labelled training set with which to develop an alert-broker for LSST, and could simply wait for the survey to begin observations. Unfortunately, no such training set exists and broker development is an open-ended research question. We developed this pipeline recognizing that it will change continuously as we prepare for LSST, with the addition of more astrophysical classes, feature characterization, classification stages and the incorporation of contextual information.

\subsection{Real Datasets as Tests of Broker Performance}

We have examined the effectiveness of machine learning algorithms as stages of an alert-broker for LSST. We developed three different stages, each addressing a different question than a generic alert-broker will have to answer -- variable-transient separation, variable and transient classification, and high-purity sample selection.

To train and test these stages, we assembled a very large labelled dataset of real variable and transient objects, including observations from a wide variety of surveys in at least two passbands. The heterogeneity and class imbalance of this dataset required that we impose additional constraints, in order to ensure that our machine-learning pipeline was making decisions based on astrophysical properties rather than characteristics of the parent surveys. These constraints are evident in the design of our feature extraction, and our choice to exclude contextual information.

Even when processing real alerts, contextual information cannot be guaranteed; the southern sky has less extant coverage, and LSST will discover sources that are more faint than any in existing surveys. Nor will such contextual information always add \emph{useful} information. Further, alert-brokers will need to test their performance on large simulated datasets of photometric time-series (e.g. PLAsTiCC) that lack such information. Simulations that do choose to include contextual information will likely only reflect our present ignorance about many of the correlations between astrophysical sources and their environment. Indeed, LSST itself will perform many of the population studies that will shed light on these relationships. 

Despite these limitations, working with real datasets has several advantages. The extreme class imbalance of our dataset presents a much more realistic challenge than more homogeneous training sets (real or simulated), which only represent a small subset of the astrophysical sources that LSST can expect to discover. Additionally, real data has several pathologies (outlying data, gaps due to bad weather, calibration errors) that are often not reflected in simulations. 

The heterogeneity of our dataset forced us to adopt more advanced feature extraction schemes, and carefully consider how we trained our machine learning pipeline in order to avoid bias. This dataset can be used to benchmark the algorithms, exposing previously unconsidered failure modes and setting an effective lower bound for machine learning performance. Each of the stages we designed for this work has benefited from the use of a real labelled dataset, especially because we can address specific use cases for an alert-broker such as \texttt{ANTARES}, as outlined in \S\ref{sec:overview}. We expect this pipeline to be even more effective on homogeneous datasets. However, the real benefit of improved datasets is that we improve the stages themselves further. Below, we discuss how each of the stages in this pipeline might evolve as we enter the era of LSST. 

\subsection{Performance of the Variable--Transient Categorization}

With only two bands, the variable vs transient classifier achieves an AUC of 0.99. This is partly due to the inclusion of the timescale uncertainty and false alarm probability along with the characteristic timescale as features (as these features are sensitive to the shape of the signal), as well as the use of timescale information from two independent passbands, which are strongly correlated in the case of recurring variables.

With the addition of contextual information from surveys such as GAIA and filtering stages such as the VPDF , we will be able to improve variable transient separation further, allowing us to perform categorization with fewer observations. After a year of LSST survey operations, we will have a long baseline for recurrent variables, effectively guaranteeing nearly perfect categorization with only a few observations of a new alert source. This will enable follow-up studies that probe the early-time evolution of several classes of astrophysical transients, giving us a window into the physics of their progenitor systems. We can further improve this capacity by including the periodogram vector itself as a feature. This will allow source classification, rather than just categorization. Together with contextual information, such a stage will be useful to rapidly identify new transients in the alert stream that are likely members of a particular astrophysical class, enabling prioritized fast-turnaround followup. 

We can also examine the use of neural networks and deep learning techniques to identify relevant features from image data, such as postage stamp cutouts included with the alerts, rather than relying solely on time-series data. This will allow the extraction of the maximum amount of information from each LSST alert packet.

\subsection{Performance of the Variable \& Transient Classifier}

Broad classification of the alert stream is the core function of any alert-broker, and indeed can be considered synonymous with brokering itself. To accomplish this task, we extracted informative astrophysical features from our heterogeneous dataset, and we developed effective techniques to address the extreme class imbalance. This allowed us to train a random forest that could successfully distinguish several broad astrophysical classes with an average accuracy of $\sim 96\%$.

However, the limitations of the dataset meant that neither the classes represented, nor the features derived, were complete. Adding other classes (e.g. AGN) would require adding other surveys, as well as other features. These features could be used to distinguish between the surveys, leading to the development of biased classifiers that cannot be adapted to new surveys.

Further development of this stage will benefit the most from a new homogeneous dataset, such as light curves from the Pan-STARRS Medium Deep Survey. However, this dataset is not labelled, and we will need to use this pipeline to begin to classify the PS1 light curves with the goal of retraining our existing classifiers and developing new stages.

Another interesting avenue for development is joint human--machine classification, where a subset of objects are provided to the public using citizen science portals such as Zooniverse\footnote{\url{https://www.zooniverse.org/}}. Human vetted objects can be used to validate the output of the brokers, as well as improve classification performance on edge cases, where visual inspection may identify features that have not been considered previously.

\subsection{Performance of the SNIa vs Non-SNIa classifier}

For the SNe stage, we explored a non-parametric approach to the problem of photometric supernova classification. The peak performance for the OSC was 0.94 AUC, based on the approach developed by L16, and 0.98 AUC for the SNPhotCC. These numbers represent high quality classification, and the AUC of 0.98 is at the same level as L16's performance when using wavelet features. It outperforms or is very competitive with other SNe classification algorithms that have been benchmarked with the SNPhotCC dataset. 

The BAGIDIS encoding of the light curve is much smaller (almost a factor of two) than the SWT coefficients, yet both methods perform equally well on the SNPhotCC data. Classification is less effective on the real OSC light curves than the simulated SNPhotCC, with the performance of BAGIDIS dropping significantly. The Unbalanced-Haar transform underlying BAGIDIS is more sensitive to sharp changes because of the shapeness of the Haar wavelet. The relative lack of stability (and thus more noisy sharp changes) for the OSC Gaussian process fits as compared to that of SNPhotCC implies that the Haar wavelet coefficients are a less reliable way to capture information from the OSC light curves. This likely reflects the heterogeneity of the OSC dataset, which contains photometry from several different telescopes and instruments, calibrated with varying degrees of photometric precision, spanning three decades. This outperformance of both wavelet methods on the SNPhotCC dataset relative to the OSC dataset bodes well for analysis of future LSST light curves. The data stream from LSST will have consistent photometry, and ultimately be more similar to the SNPhotCC dataset than the heterogeneous OSC. The AUC value of 0.98 for both wavelet schemes is very competitive, and highlights the success of the non-parametric approach to light curve classification.

The biggest improvements to this stage will be the addition of more independent feature extraction methods and more astrophysical classes in the training. All classification with a large number of observations is retrospective, and there are few additional observations that can be obtained that will add useful new information. Consequently, when considering the full time-evolution and contextual information for sources, brokers must be able to extract extremely high purity samples for \emph{any well-defined} astrophysical class. Brokers that can surmount this high bar will be able to produce ``living'' catalogs of the variable and transient sky, extracting the maximum gain from the alert stream. 

\section{Future Work}\label{sec:futurework}

The fundamental challenge for \texttt{ANTARES} and all alert-brokers is that no present survey can generate even a limited number of alerts with similar properties as LSST. The extant labelled datasets suitable for supervised learning algorithms are small, and drawn from a mixture of different surveys. Simulated datasets, while useful, often do not accurately reflect the pathologies of real data. Many correlations that are present in real data, such as those between transients and their host environment, or variables and their location in the galaxy, are simply not captured by simulations. 

While unlabelled data exist from sources such as PS1 Medium Deep Survey data and DES, much of the data to classify them effectively, such as publicly available detection tables, do not exist. The most promising approach then is semi-supervised learning: where a smaller training set is used to classify a fraction of a larger unlabelled set, which is then in turn used to retrain the classifier. We plan to perform a retrospective classification of light curves from the Pan-STARRS PS1 medium deep survey, which we will treat as a semi-supervised learning problem. We also welcome the contribution of models or data from the astronomical community, and we are keen to develop new stages that meet the needs of the many groups involved in time-domain science. 

One of the most pressing developments for \texttt{ANTARES} will be adding a ``None of the above'' label, and consistent criteria for a source to meet to be flagged as such. Identifying and potentially clustering outliers that do not belong to any previously known astrophysical class is the core focus of \texttt{ANTARES}. The stages that result from these studies will be tested extensively by the upcoming PLAsTiCC dataset.

Significant work will also be devoted to the front-end interface for \texttt{ANTARES}, which we will begin to test publicly. We will begin running on various live alert streams from surveys, including the ASAS-SN and upcoming ZTF public survey alerts. These studies in particular are essential for developing and battle-testing \texttt{ANTARES} before the commencement of the LSST survey. 

\acknowledgments

GN is supported by the Lasker Fellowship at the Space Telescope Science Institute. TZ began working on \texttt{ANTARES} under the NOAO Summer REU program (2015, adviser: GN). TZ elected to continue working with GN on his senior honors thesis on supernova light curve classification at Macalester College (earned April 2017). Material from TZ's thesis is presented in this work. ML acknowledges support from the SKA SA, the NRF and AIMS. \texttt{ANTARES} is generously supported by NSF INSPIRE grant (CISE AST-1344204, PI: R. Snodgrass) as well as funding to develop the NOAO Community Science Data Center (CSDC).

This work makes extensive use of the OGLE catalog and the Open Supernova Catalog, and we are grateful to both projects for maintaining open-access portals to data. We are very grateful to James Guillochon for his help with ingesting the OSC data into \texttt{ANTARES}. This work makes extensive use of open source software and high performance computing resources, and would not be possible without the diligent efforts of the system administrators of the \texttt{ANTARES} cluster, Adam Scott and Pete Wargo, as well as the IT staff at NOAO, STScI and the University of Arizona. The authors thank Amali Vaz for her careful reading and useful suggestions. We are grateful to the anonymous referee for their detailed feedback, which has helped refine this work.
This work made extensive use of several open source packages including: 
\software{\texttt{numpy, scipy, matplotlib, astropy, pandas, pyMySQL, H5py, pyFFTW, gatspy, george, PyWavelets, imbalanced-learn, scikit-learn, SLURM, MPI4py}} and the \texttt{conda} package and environment management systems.

\bibliographystyle{aasjournal} 
\bibliography{Bibliography} 

\clearpage
\end{document}